\documentclass[preprint,floats,aps, superscriptaddress,nofootinbib,floatfix,showkeys]{revtex4-2}
\usepackage{natbib}
\usepackage{times} 
\usepackage{amssymb,amsmath}
\usepackage{csquotes}
\usepackage{physics}
\usepackage{bm}
\usepackage{nicefrac}
\usepackage{mathtools}
\usepackage{setspace}
\usepackage{booktabs}
\usepackage{upgreek}
\usepackage[per-mode=fraction]{siunitx}

\usepackage[usenames,dvipsnames]{color}
\usepackage[pdftex]{epstopdf}
\usepackage[pdftex]{graphicx}
\usepackage{hyperref}
\hypersetup{pdftitle={Manuscript on } %ADAPT TEXT 
    pdfauthor={Florian Voss and Uwe Thiele},  %ADAPT TEXT 
pdfproducer={lateX},
pdfview=FitV,       % FitH
pdfstartview=FitB,
linkcolor=blue,     % links to same page
citecolor=blue,     % citations
urlcolor=blue,      % links to URLs
breaklinks=true,    % links may be split onto 2 lines
colorlinks=true,
citebordercolor=0 0 0,  % color for \cite
filebordercolor=0 0 0,
linkbordercolor=0 0 0,
menubordercolor=0 0 0,
urlbordercolor=0 0 0,
pdfhighlight=/I,
pdfborder=0 0 0,   % no box around links
bookmarksopen=true,
bookmarksnumbered=true
}
\usepackage{tcolorbox}
\DeclareGraphicsExtensions{.jpg, .pdf, .tif, .png}
\usepackage[normalem]{ulem}
\definecolor{reviewchangecolor}{rgb}{1, 0, 0}

\newcommand{\subref}[2]{\hyperref[#1]{\ref{#1}#2}}

\newcommand{\vecg}[1]{\boldsymbol{#1}}
\newcommand{\tens}[1]{\mathbf{\underline{#1}}}

\usepackage{booktabs}
\usepackage{multirow}
\makeatletter
\renewcommand{\p@subsection}{}
\renewcommand{\p@subsubsection}{}
\renewcommand{\@seccntformat}[1]{\csname the#1\endcsname\quad}
\makeatother

\begin{document}
\count\footins = 1000 % fixes behavior with long footnotes

\title{Chemomechanical motility modes of partially wetting liquid droplets}

\author{Florian Voss}
\email{\textcolor{blue}{f\_voss09@uni-muenster.de}}
\thanks{ORCID ID: 0009-0003-9679-035X}
\affiliation{Institute of Theoretical Physics, University of M\"unster, Wilhelm-Klemm-Str.\ 9, 48149 M\"unster, Germany}

\author{Uwe Thiele}
\email{\textcolor{blue}{u.thiele@uni-muenster.de}}
\homepage{http://www.uwethiele.de}
\thanks{ORCID ID: 0000-0001-7989-9271}
\affiliation{Institute of Theoretical Physics, University of M\"unster, Wilhelm-Klemm-Str.\ 9, 48149 M\"unster, Germany}
\affiliation{Center for Nonlinear Science (CeNoS), University of M\"unster, Corrensstr.\ 2, 48149 M\"unster, Germany}
\affiliation{Center for Multiscale Theory and Computation (CMTC), University of M\"unster, Corrensstr.\ 40, 48149 M\"unster, Germany}

\begin{tcolorbox}[title=Publication note,
title filled=false,
colback=red!5!white,
colframe=red!75!black,sharp corners]
This version of the article has been accepted for publication after peer-review. It has been modified to account for most post-acceptance improvements and corrections, but it is not the version of record. The version of record is available at:
\\
\\
F.~Voss and U.~Thiele.
\newblock Chemomechanical motility modes of partially wetting liquid droplets.
\newblock \emph{Phys. Rev. Fluids}, 10:\penalty0 094005, 2025{\natexlab{a}}.
\newblock \doi{10.1103/f3ck-dx5c}.
\end{tcolorbox}

\begin{abstract}
We consider a simple thermodynamically consistent model that captures the self-organized chemomechanical coupling resulting from the interplay between autocatalytically reacting surfactants, the Marangoni effect and wetting dynamics. An ambient bath of surfactant acts as a chemostat and provides the system with chemical fuel, thereby driving it away from thermodynamic equilibrium. We find that a positive feedback loop between the local reactions and the Marangoni effect induces surface tension gradients that allow for self-propelled droplets. Besides simple directional motion, we find crawling and shuttling droplets as well as droplets performing random walks, thus exploring the entire substrate. We study the occurring chemomechanical motility modes and show how the observed dynamic states emerge from local and global bifurcations. Due to the underlying generic thermodynamic structure, we expect that our results are relevant not only to directly related biomimetic droplet systems but also to structurally similar systems like chemically active phase-separating mixtures.
\end{abstract}

\maketitle
\section{Introduction}
The interface between hydrodynamics and chemistry is rich in fascinating phenomena, ranging from chemical gardens \cite{BCCC2015cr} and chemically driven active colloids \cite{GoLA2005prl} to periodically erupting droplets \cite{SYM2023sr}. Considering this ubiquitous complexity, it seems plausible that also many biological systems are found at the intersection of these fields \cite{GrKG2017arb,MSGE2019prf,RaTh2004arb, MJRL2013rmp}. Correspondingly, the study of physical processes in living matter, including hydrodynamic phenomena, has gained considerable traction in recent years. Examples include hydrodynamic models of the actomyosin complex \cite{JKPJ2007pr,BoJG2011prl,YoBG2016pd}, mitotic spindle positioning \cite{Shel2016arfm}, chromatin dynamics \cite{BGRZ2014bj}, protoplasmic droplets \citep{RaEB2014po,KuJBE2019po} and osmotic biofilm spreading \cite{SAWV2012pnasusa, TJLT2017prl} as well as studies of reaction- and diffusion-based protein dynamics at biomembranes \cite{JoBa2005pb,JoBa2005prl,HaFr2018np,HaBF2018ptrsbs} and biomolecular condensates \cite{BLHR2017nrmcb, HyWJ2014arcdb,KiZw2021jotrsi,DGMF2023prl, GDMF2024prr}. Here, minimalistic models complement complicated biologically faithful descriptions, as they are more accessible to theoretical and conceptual study. In particular, motility and self-propulsion phenomena are studied as hallmarks of active matter, with diverse underlying physico-chemical mechanisms including asymmetric distributions of catalytic activity \cite{ISBW2002acie,PKOS2004jotacs, GoLA2005prl, HJRG2007prla, GoLA2007njop, MiLa2017sr}, self-induced wettability gradients \cite{DoOn1995prl,LeLa2000jacs,LeKL2002pre, SKYN2005pre, ThJB2004prl, JoBT2005epje}, active stresses \cite{RaEB2014po,KuJBE2019po,ZiSA2012jrsi, TTMC2015nc,TSJT2020prea, StJT2022sm} and enzymatically maintained concentration gradients \cite{DGMF2023prl, GDMF2024prr}. Among such self-propulsion strategies, the (solutal) Marangoni effect,~i.e., the emergence of mechanical forces localized at interfaces that results from surface tension gradients, is especially common in the context of biomimetic and prebiotic systems. It is employed in,~e.g., drop-based microswimmers \cite{MKHB2016arcmp,Mich2023arfm} and some simple models of protocell motility \cite{MaNK2023c}. Aimed at achieving biomimetic functions, previous studies have combined autocatalytic pattern-forming reactions like the Belousov-Zhabotinsky reaction with (droplet) hydrodynamics \cite{KAMY2002jcp,KYNS2011pre, SMAN2016jpcl,SzGH2013jpcc} and polymeric gels \citep{LuRGZW2013cc, RYG2020sa, YaBa2006s} resulting in complex deformation and motility modes due to chemically driven flows.  Autocatalytic mechanisms are also discussed as possible forms of molecular self-replication under prebiotic conditions \cite{PaJo2004cocb, RuBE2013cr,AESK2020nrc}. Interestingly, candidates for autocatalytic self-replication also include,~e.g., amphiphilic peptides \cite{RWRA2009ac} that can adsorb at water-air interfaces. In view of these observations, it is compelling to study whether the coupling of autocatalytic processes at interfaces and droplet hydrodynamics can result in complex forms of motility without the highly specialized biochemical machinery of biological cells. 

Here, we propose a simple mesoscopic hydrodynamic model that captures the interplay of an autocatalytic reaction at the free surface of a droplet, the solutal Marangoni effect and the wetting dynamics in the presence of chemical fuel. Despite being conceptually simple, the model exhibits a striking degree of complexity of the resulting modes of self-propulsion. 

We remark that the model briefly considered in Sec.~4.2 of Ref.~\citep{VoTh2024jem} by the authors is related to the one presented in this article. A notable difference consists in the modeling of the exchange reactions with the chemostats (which provide the chemical fuel) which are here, in contrast to Ref.~\citep{VoTh2024jem}, not linearized about chemical equilibrium. More importantly, here, we consider parameter regimes for which the reaction-diffusion subsystem of surfactants without any coupling to hydrodynamics does not form any patterns. Instead, the various modes of droplet motion emerge only as a consequence of a chemomechanical feedback loop that results from the interplay of nonlinear chemical reactions and the Marangoni effect which we describe in Sec.~\ref{sec:simple_traveling}.

The present article is organized as follows. In Sec.~\ref{sec:model} we provide a thermodynamically consistent free-energy based description of droplets covered by chemically reacting surfactants. We discuss how the presence of chemical fuel results in persistent nonequilibrium behavior. In Sec.~\ref{sec:simple_traveling} we first investigate droplets on a one-dimensional substrate and study the underlying self-propulsion mechanism which we identify as a positive feedback loop between the local chemical reactions and the Marangoni effect. In Sec.~\ref{sec:bifurcation_analysis} we study more complex forms of droplet motion, namely periodic stick-slip-like motion (\enquote{crawling}) and back-and-forth motion (\enquote{shuttling}), and the related bifurcations. Finally, we briefly turn to droplets on a two-dimensional substrate, where the additional degree of freedom allows for highly complex types of motility. In Sec.~\ref{sec:conclusion}, we recapitulate our findings, discuss potential implications for biomimetic and related biological or prebiotic systems and list possible extensions of the model.

\section{Droplets covered by autocatalytic surfactants}\label{sec:model}
\begin{figure}
\centering
\includegraphics[scale=0.63]{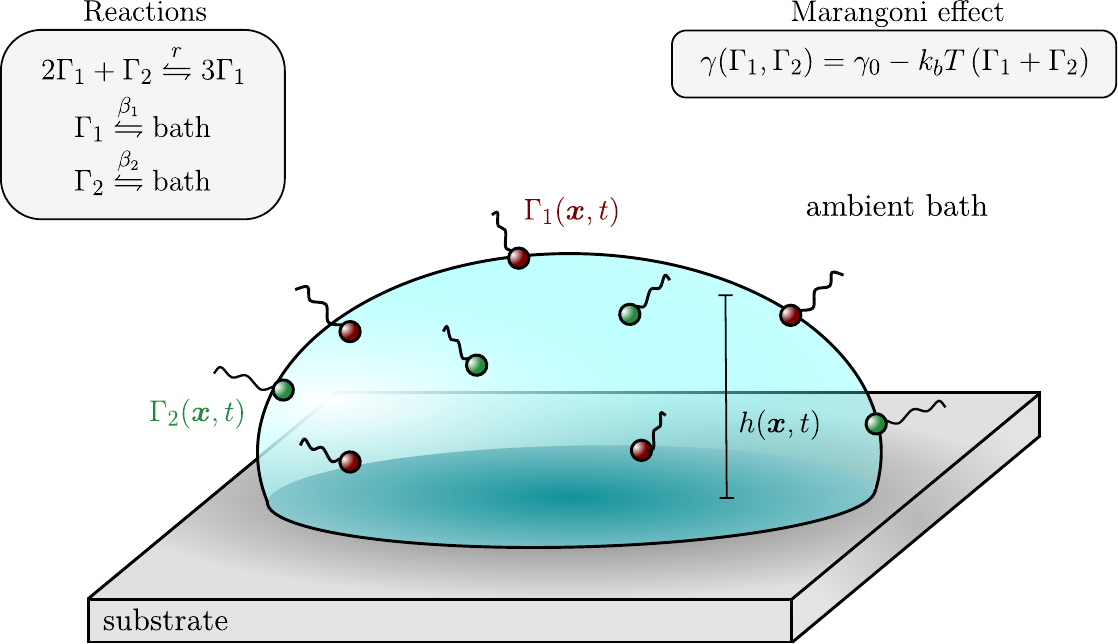}
\caption{Sketch of the considered system. A droplet of a partially wetting liquid is situated on a flat solid substrate. The local film thickness is denoted by $h(\vecg{x},t)$. The droplet is in contact with an ambient bath and its free surface is populated by two species of surface-active molecules (surfactants) with densities $\Gamma_1(\vecg{x},t)$ and $\Gamma_2(\vecg{x},t)$. They equally linearly reduce the local surface tension $\gamma$ and engage in an autocatalytic reaction with reaction rate $r>0$. Here, $\Gamma_1$ catalyses its own production. The ambient bath acts as a surfactant source or sink (chemostat) with the exchange rates $\beta_1, \beta_2>0$.}
\label{fig:system_sketch}
\end{figure}
We consider a mesoscopic droplet of a simple incompressible, partially wetting liquid that is situated on a flat, solid substrate (Fig.~\ref{fig:system_sketch}). The free surface can be parametrized by the local film thickness $h(\vecg{x}, t)$ with substrate coordinates $\vecg{x}=(x,y)^\text{T}$ and time $t$. It is populated by insoluble surface-active chemical species (surfactants) with densities $\Gamma_1(\vecg{x}, t)$ and $\Gamma_2(\vecg{x}, t)$ (particles per unit surface area). The droplet is embedded in an ambient fluid that acts as a chemostat for surfactants. The free energy of the system is
\begin{equation}
F=\int_\mathcal{S}\left[f(h)+\xi g(\Gamma_1,\Gamma_2)\right]\text{d}^2x, \label{eq:free_energy_definition}
\end{equation}
where $\mathcal{S}$ denotes the substrate plane, $\xi=\sqrt{1+\vert\vert \nabla h\vert\vert^2}$ is the local metric factor of the droplet surface with $\nabla=(\partial_x, \partial_y)^\text{T}$, and $\vert\vert \cdot\vert\vert$ is the Euclidean norm. The partial derivative with respect to $i$ is denoted by $\partial_i$. Equation (\ref{eq:free_energy_definition}) comprises two contributions, namely the mesocopic wetting energy $f(h)$ that encodes all liquid-substrate interactions and the surface energy $g(\Gamma_1, \Gamma_2)$. We here neglect possible surfactant-substrate interactions \cite{TSTJ2018l}, such that $f$ is independent of $\Gamma_1$ and $\Gamma_2$. Conversely, $g$ is assumed to be independent of $h$. Specifically, we assume that the wetting energy is a simple superposition of long-range attractive and short-range repulsive interactions,
\begin{equation}
f(h) = A\left(-\frac{1}{2h^2}+\frac{h_a^3}{5h^5}\right). \label{eq:wetting_energy_definition}
\end{equation}
Here, $A$ is the Hamaker constant that is directly related to the equilibrium contact angle \cite{BEIM2009rmp}, and $h_a$ is the thickness of the ultrathin adsorption layer that covers the macroscopically \enquote{dry} substrate. Note that with $A>0$ we have for the spreading parameter $S=f(h_a)<0$ and Eq.~(\ref{eq:wetting_energy_definition}) corresponds to partial wetting \cite{Genn1985rmp, BEIM2009rmp,Thie2010jpcm}. We further assume that the droplet surface is only sparsely covered by surfactant. Then, interactions between individual surfactant molecules are negligible and the surfactant-dependent part of the surface energy $g$ only comprises entropic contributions, 
\begin{equation}
g(\Gamma_1,\Gamma_2) = \gamma_0+k_b T\Gamma_1\left[\ln(\Gamma_1a_1^2)-1\right]+k_b T\Gamma_2\left[\ln(\Gamma_2a_2^2)-1\right], \label{eq:surface_energy_definition}
\end{equation}
where $\gamma_0$ is the surface tension of the bare droplet surface, $a_1,a_2$ are typical surfactant length scales, $T$ is the temperature and $k_b$ is the Boltzmann constant. This choice results in the linear equation of state,
\begin{equation}
\gamma\left(\Gamma_1,\Gamma_2\right)=\gamma_0-k_b T(\Gamma_1+\Gamma_2), \label{eq:linear_Marangoni}
\end{equation}
where the surface tension $\gamma$ symmetrically depends on both species, for details see \cite{ThAP2012pf, VoTh2024jem}. This implies that only the total surfactant count determines the surface tension and therefore excludes (self-propulsion) effects based on differences in surfactant properties \cite{MKHB2016arcmp,Mich2023arfm}. The surfactants chemically react in the reversible autocatalytic reaction
\begin{equation}
2\Gamma_1+\Gamma_2 \stackrel{r}{\rightleftarrows} 3\Gamma_1, \label{eq:reaction_autocat}
\end{equation}
where $\Gamma_1$ catalyzes its own production with the reaction rate $r>0$. Consequently, we refer to $\Gamma_1$ as the autocatalyst and to $\Gamma_2$ as the reactant. Reaction (\ref{eq:reaction_autocat}) corresponds to the central nonlinear reactions of paradigmatic pattern forming systems like the Brusselator \citep{PrLe1968jcp,Nicolis1999} and the Schl\"ogl model \citep{Nicolis1999}. Here, it represents a simple incorporation of a chemical nonlinearity. In systems containing multiple surfactants, such higher order reactions may be associated with the formation of micelles that consist of both species \citep{SSK2017sm,TaNN2021sm}. The droplet also exchanges surfactant with the ambient bath which acts as a reservoir or chemostat for both chemical species. This is modeled as reversible adsorption-desorption reactions,
\begin{equation}
\begin{aligned}
\Gamma_1 &\stackrel{\beta_1}{\rightleftarrows}  \text{bath of chemical potential $\mu_1$},\\
\Gamma_2 &\stackrel{\beta_2}{\rightleftarrows}  \text{bath of chemical potential $\mu_2$},
\end{aligned}\label{eq:reservoir_exchange}
\end{equation}
where forward reactions (e.g.,~$\Gamma_1\rightarrow\text{bath}$) correspond to desorption from the free surface and $\beta_1, \beta_2>0$ are the reaction rates for $\Gamma_1$ and $\Gamma_2$, respectively. Instead of treating an ambient bath, one could equivalently assume an exchange with the droplet bulk as a reservoir. In both cases, we assume that the reservoir is large such that the exchange of surfactant with the drop surface does not significantly alter the concentrations of species in the bath and the corresponding constant chemical potentials $\mu_1$ and $\mu_2$ are convenient control parameters. For a thermodynamically sound justification of this approximation, see Ref.~\citep{ReEA2025jcp}. The reactions (\ref{eq:reservoir_exchange}) can therefore continuously provide the droplet with chemical fuel and drive the system away from thermodynamic equilibrium. The dynamics is then described by a \enquote{passive core} in gradient dynamics form with thermodynamic forces derived from variations of the free energy (\ref{eq:free_energy_definition}), that is augmented by the exchange reactions with the bath. The complete model reads
\begin{alignat}{3}
	&\partial_t h &&= -\nabla\cdot\bm{j}_h &&= \nabla\cdot\left[Q_{hh}\nabla\frac{\delta F}{\delta h}+Q_{h\Gamma_1}\nabla\frac{\delta F}{\delta \hat{\Gamma}_1}+Q_{h\Gamma_2}\nabla\frac{\delta F}{\delta \hat{\Gamma}_2}\right]\nonumber,\\
	&\partial_t \hat{\Gamma}_1 &&= -\nabla\cdot\bm{j}_1 +\mathcal{R}+\mathcal{B}_1 &&= \nabla\cdot\left[Q_{\Gamma_1h}\nabla\frac{\delta F}{\delta h}+Q_{\Gamma_1\Gamma_1}\nabla\frac{\delta F}{\delta \hat{\Gamma}_1}+Q_{\Gamma_1\Gamma_2}\nabla\frac{\delta F}{\delta \hat{\Gamma}_2}\right]+\mathcal{R}+\mathcal{B}_1\label{eq:model},\\
	&\partial_t \hat{\Gamma}_2 &&= -\nabla\cdot\bm{j}_2 -\mathcal{R}+\mathcal{B}_2 &&= \nabla\cdot\left[Q_{\Gamma_2h}\nabla\frac{\delta F}{\delta h}+Q_{\Gamma_2\Gamma_1}\nabla\frac{\delta F}{\delta \hat{\Gamma}_1}+Q_{\Gamma_2\Gamma_2}\nabla\frac{\delta F}{\delta \hat{\Gamma}_2}\right]-\mathcal{R}+\mathcal{B}_2.\nonumber
\end{alignat}
Here, $\hat{\Gamma}_{1,2}=\xi\Gamma_{1,2}$ are \enquote{projected} densities (particles per unit substrate area) that directly correspond to the particle numbers of surfactant and are \textit{independent} of the surface geometry, for details see \cite{ThAP2012pf, ThAP2016prf, VoTh2024jem}. For a general account of the thermodynamic structure of Eqs.~(\ref{eq:model}) we also refer to \cite{VoTh2024jem} and references therein. The fully equivalent hydrodynamic form is given in Appendix \ref{app:hydrodynamic_form}. Note that Eqs.~(\ref{eq:model}) conserve the total liquid volume. The variational derivatives $\delta F/\delta h$ and $\delta F/\delta \hat{\Gamma}_{1}, \delta F/\delta hat{\Gamma}_{2}$ correspond to the liquid pressure and the chemical potentials of $\Gamma_{1}$ and $\Gamma_{2}$ on the droplet surface, respectively. The transport fluxes $\bm{j}_h$ and $\bm{j}_1,\bm{j}_2$ in Eqs.~(\ref{eq:model}) are thus linear in gradients of pressure and chemical potentials and represent diffusive and advective contributions to transport, including Marangoni fluxes. The associated mobility matrix \cite{ThAP2012pf, VoTh2024jem},
\begin{equation}
	\tens{Q} = \left(\begin{array}{ccc}
Q_{hh} & Q_{h\Gamma_1} & Q_{h\Gamma_2} \\ 
Q_{\Gamma_1h} & Q_{\Gamma_1\Gamma_1} & Q_{\Gamma_1\Gamma_2} \\ 
Q_{\Gamma_2h} & Q_{\Gamma_2\Gamma_1} & Q_{\Gamma_2\Gamma_2}
\end{array} \right) = \left(\begin{array}{ccc}
	\frac{h^3}{3\eta} & \frac{h^2\Gamma_1}{2\eta} & \frac{h^2\Gamma_2}{2\eta} \\ 
	\frac{h^2\Gamma_1}{2\eta} & \frac{h\Gamma_1^2}{\eta}+D_1\Gamma_1 & \frac{h\Gamma_1\Gamma_2}{\eta}\\ 
	\frac{h^2\Gamma_2}{2\eta} &  \frac{h\Gamma_1\Gamma_2}{\eta}  &  \frac{h\Gamma_2^2}{\eta}+D_2\Gamma_2
	\end{array} \right),\label{eq:autocatalyic_surfactants_Q}
\end{equation}
is positive definite and symmetric, ensuring non-negative entropy production and satisfying the Onsager relations \cite{GrootMazur1984,Onsa1931prb,Onsa1931pr}. It corresponds to a thin-film description for droplets covered by insoluble surfactants without slip at the substrate \cite{ThAP2012pf, VoTh2024jem}. Here, $\eta>0$ is the dynamic viscosity of the liquid and $D_1, D_2>0$ are diffusive mobilities of the respective surfactant. Note that in the underlying hydrodynamic problem, the transport fluxes $\bm{j}_h$ and $\bm{j}_1,\bm{j}_2$ respectively correspond to the vertically integrated horizontal components of the liquid velocity and to diffusive and advective surfactant transport at the free surface \citep{PTTK2007pf}. Equation (\ref{eq:autocatalyic_surfactants_Q}) implies that the ambient medium does not contribute to the drop dynamics, e.g., because it is of low viscosity. This differs from the commonly treated Marangoni-driven microswimmers \cite{MKHB2016arcmp,Mich2023arfm}. However, this assumption could be relaxed in the future by incorporating aspects of \citep{MPBT2005pf}. The model (\ref{eq:model}) could also be extended to treat soluble surfactants \citep{ThAP2016prf}, e.g., to capture a possible depletion or enrichment of surfactants in the bath. Then, the chemical potentials $\mu_1, \mu_2$ would be dynamical quantities, instead of the explicit chemostatting assumed here.

Unlike the transport fluxes, the autocatalytic current $\mathcal{R}$ is nonlinear in the free energy variations,
\begin{equation}
\mathcal{R}=r\left[\exp\left(\frac{2}{k_b T}\frac{\delta F}{\delta \hat{\Gamma}_1}+\frac{1}{k_b T}\frac{\delta F}{\delta \hat{\Gamma}_2}\right)-\exp\left(\frac{3}{k_b T}\frac{\delta F}{\delta \hat{\Gamma}_1}\right)\right] \label{eq:autocatalysis_detailed_balanced},
\end{equation}
where the structure of Eq.~(\ref{eq:autocatalysis_detailed_balanced}) expresses the principle of detailed balance \cite{GrootMazur1984,WZJL2019rpp,Zwic2022cocis,VoTh2024jem} and contains the flux of the forward and the backward reactions of Eq.~(\ref{eq:reaction_autocat}) as the first and second term, respectively. In the limit of ideal systems, Eq.~(\ref{eq:autocatalysis_detailed_balanced}) corresponds to standard mass-action kinetics. So far, all contributions are purely passive,~i.e., they result in a decrease of the free energy $F$ until thermodynamic equilibrium is attained. 

The final nonconserved terms in Eqs.~(\ref{eq:model}) model the exchange of surfactant with the ambient bath and are similarly to Eq.~(\ref{eq:autocatalysis_detailed_balanced}) given by
\begin{equation}
\begin{aligned}
\mathcal{B}_1&=\beta_{1}\left[\exp\left(\frac{\mu_1}{k_b T}\right)-\exp\left(\frac{1}{k_b T}\frac{\delta F}{\delta \hat{\Gamma}_{1}}\right)\right],\\
\mathcal{B}_2&=\beta_{2}\left[\exp\left(\frac{\mu_2}{k_b T}\right)-\exp\left(\frac{1}{k_b T}\frac{\delta F}{\delta \hat{\Gamma}_{2}}\right)\right],
\end{aligned}
\label{eq:adsorption_desorption}
\end{equation}
where $\mu_1, \mu_2$ are the uniform and constant chemical potentials of $\Gamma_1$ and $\Gamma_2$ in the bath that acts as a chemostat. This reflects that the concentrations in the bath are not significantly affected by the exchange with the drop surface. Note that in Sec.~4.2 of Ref.~\citep{VoTh2024jem}, the linearized expressions of Eqs.~(\ref{eq:adsorption_desorption}) are employed, which are strictly valid only near thermodynamic equilibrium \citep{GrootMazur1984}. For ${\mu_1=\mu_2}$, Eqs.~(\ref{eq:model}) represent an open albeit passive system where the grand potential $F-\int\sum_i\xi\mu_i\Gamma_i\text{d}^2x$ continuously decreases.\footnote{See,~e.g., the formulation of the second law of thermodynamics for open reaction-diffusion systems in Ref.~\citep{AvAFE2024} which also applies here.} Then, the system ultimately relaxes to a single droplet that is uniformly covered by both surfactants. Otherwise, $\mu_{1}$ and $\mu_{2}$ reflect driving forces and the system is permanently out-of-equilibrium. More generally, the nonequilibrium driving results from an incompatibility between the chemical potentials of the chemostat with respect to the conditions for thermodynamic equilibrium \citep{AvAFE2024}. Thus, if the chemical potentials of the chemostat are uniform and constant, two or more chemostatted species are necessary for sustained nonequilibrium. Therefore, systems with only one chemostat such as the chemically driven running drop discussed in Sec.~4.1 of Ref.~\citep{VoTh2024jem} indeed relax toward thermodynamic equilibrium, i.e.\ the (semi-)grand potential continuously decreases \citep{AvAFE2024} (in contrast to the increasing free energy shown in Ref.~\citep{VoTh2024jem}). However, if the running drop moves in an infinite domain the actual equilibrium is never reached somewhat similar to a drop sliding down an infinite incline.

To study the emerging dynamics, we nondimensionalize Eqs.~(\ref{eq:model}). Importantly, we explicitly use the assumption that slopes in the droplet profile are small ($\vert\vert\nabla h\vert\vert\ll 1$), resulting in the approximation $\hat{\Gamma}_{1,2}\approx\Gamma_{1,2}$ in the final equations. For details, we refer to Appendix \ref{app:nondimensionalization}, where we also discuss possible dimensional parameter values. From hereon, all quantities are nondimensional. In the following, we investigate the dynamics of self-propelled droplets which occur for sufficiently strong nonequilibrium driving.

\section{Self-propulsion mechanism}\label{sec:simple_traveling}

\begin{figure}
\centering
\includegraphics[scale=1]{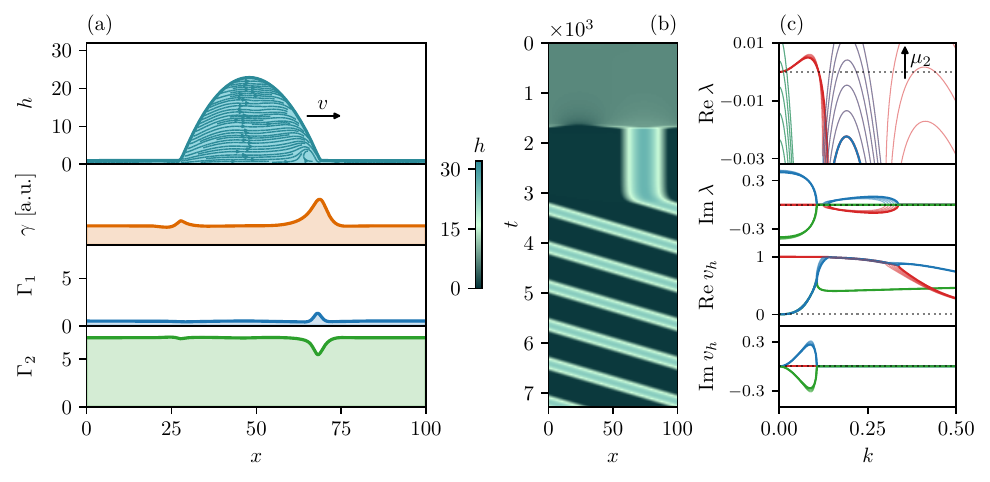}
\caption{Flat films rupture via spinodal dewetting and self-organize into self-propelled droplets moving with constant velocity $v$. (a) shows the final self-propelled state, where the droplet moves across the substrate driven by a net imbalance in surface tension between the front and the rear. The top panel shows the film thickness profile $h$, the other panels show the profiles of the surface tension $\gamma$ and the surfactant concentrations $\Gamma_1$ and $\Gamma_2$. The streamlines in the top panel correspond to the velocity field of the liquid in the laboratory frame. Panel (b) shows a space-time diagram of the initial dewetting process and subsequent self-propulsion. (c) Results of the linear stability analysis of a flat film that corresponds to the initial condition of (b). The top two panels give the real and imaginary parts of the eigenvalues $\lambda$ (red, blue, green) as functions of the wavenumber $k$. The bottom two panels show the real and imaginary parts of the $h$-component $v_h$ of the three normalized eigenvectors as functions of $k$ (also red, blue, green). The dotted lines indicate zero. The flat film is linearly unstable for small wavenumbers (red) corresponding to a Cahn-Hilliard (spinodal) instability which does not couple to the surfactant fields. When $\mu_2$ is increased (thin lines), other wavenumbers become unstable,~e.g., via a Hopf-instability (top panel, green line crosses zero at $k=0$), via a wave instability (top panel, blue line crosses zero at $k\neq 0$) or via a Turing instability (top panel, red line crosses zero at $k\neq 0$). The parameters for (a),~(b),~(c) are $\mu_1=-1.4, \mu_2=4.13, r=0.3, \beta_1=2, \beta_2=0.01, \delta=1, D_1=D_2=0.45, W=10$ with a mean film thickness of $\bar{h}=7$. The computational domain in (a),~(b) is [0,100]. In (c), the thin lines correspond to $\mu_2=4.15,4.17,4.19,4.21,4.23$. See also Supplemental Video 1.}
\label{fig:selfpropelled_droplets_overview}
\end{figure}
\begin{figure}
\centering
\includegraphics[scale=1]{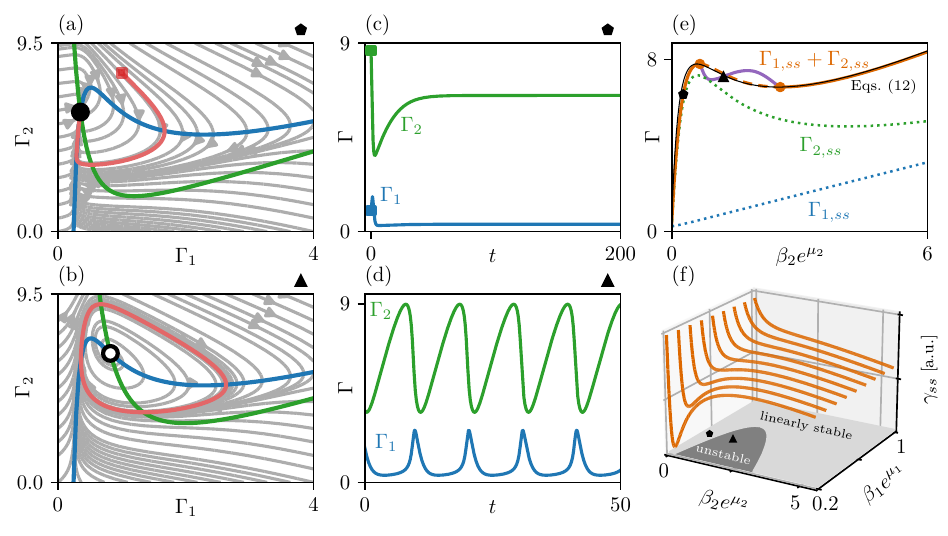}
\caption{Properties of the local reactor (\ref{eq:local_reactor}). Panels (a) and (b) show the phase portraits outside and inside the Hopf unstable region with $\mu_2=3.3$ and $\mu_2=4.8$, respectively. The nullcline of $\Gamma_1$ ($\Gamma_2$) is represented as a blue (green) line. The stable (unstable) fixed point is shown as a filled (empty) circle symbol. Red lines represent typical trajectories, in (a) the red square denotes the initial condition, in (b) the trajectory corresponds to the limit cycle. Panels (c) and (d) show the time evolution of $\Gamma_1$ (blue) and $\Gamma_2$ (green) for the red trajectories in (a) and (b), respectively. In (c), the squares denote the initial concentrations. Panel (e) shows the numerically computed bifurcation diagram with $\beta_2e^{\mu_2}$ as control parameter ($\beta_2$ fixed). The steady state concentrations $\Gamma_{1,ss}$ and $\Gamma_{2,ss}$ are drawn in blue and green (dotted lines) the sum of both concentrations is shown in orange [black for estimate given by Eqs.~(\ref{eq:reactor_fp})]. Supercritical Hopf bifurcations occur at the extrema of $\Gamma_{1,ss}+\Gamma_{2,ss}$ (orange dots). Between them, the steady state is unstable (dashed line). In the unstable region, a limit cycle exists, where the mean total concentration is represented as a purple solid line. (f) Dependence of the steady state surface tension $\gamma(\Gamma_{1,ss}+\Gamma_{2,ss})$ on both driving currents $\beta_1e^{\mu_1},\beta_2e^{\mu_2}$ (fixed $\beta_1,\beta_2$) given by Eqs.~(\ref{eq:linear_Marangoni}) and (\ref{eq:reactor_fp}). The stability diagram shown at the bottom results from Eq.~(\ref{eq:reactor_hopf_bifs}). In Panels (e) and (f), pentagon [triangle] markers denote the parameter choices in (a) and (c) [(b) and (d)]. The remaining parameters for all panels are $r=0.3, \beta_1=2, \beta_2=0.01, \delta=1$ and [except (f)] $\mu_1=-1.4$.}
\label{fig:ode_summary}
\end{figure}
First, we consider 2D droplets (liquid ridges) on a one-dimensional domain with periodic boundary conditions, and study the dynamics using finite-element based time simulations implemented in oomph-lib \cite{HeHa2006} (Appendix \ref{app:numerical_methods}). We specifically examine scenarios of strong molecular interactions between the liquid and the substrate as compared to the energetic influence of surfactant ($W=\frac{Aa_1a_2}{h_a^2k_b T}\gg 1$, Appendix \ref{app:nondimensionalization}). In both the passive and the active case, flat films then typically rupture by spinodal dewetting (i.e., by a long-wave instability of Cahn-Hilliard type, see Supplemental Material of \cite{FrTh2023prl} for a recent classification) and quickly form a single droplet [Fig.~\subref{fig:selfpropelled_droplets_overview}{(b)} and Fig.~\subref{fig:selfpropelled_droplets_overview}{(c)}]. At large driving force also other instability types can occur [Fig.~\subref{fig:selfpropelled_droplets_overview}{(c)}], although here we focus on the spinodal scenario. If the ambient reservoir is depleted of the autocatalyst $\Gamma_1$ and rich in the reactant $\Gamma_2$ ($\mu_1 <0, \mu_2>0$), droplets formed by dewetting spontaneously break their left-right symmetry and move across the substrate [Fig.~\subref{fig:selfpropelled_droplets_overview}{(a)} and Fig.~\subref{fig:selfpropelled_droplets_overview}{(b)}, Supplemental Video 1]. This is due to Marangoni convection induced by a greatly increased surface tension in the advancing contact line region. The surface tension in the receding contact line region is also slightly increased. The effect at the receding contact line vanishes when the coupling of the liquid pressure to the surfactant dynamics [see Eq.~(\ref{eq:autocatalyic_surfactants_Q})] is neglected and results from pressure gradients in the contact line regions. Because this effect is comparatively small we do not discuss it further.

Importantly, self-propelled droplets can be observed even when both surfactants diffuse equally ($D_1\!=\!D_2$) which excludes a Turing instability of the reaction-diffusion subsystem as an underlying mechanism. We now show that the mechanism for generating and maintaining local gradients in surface tension arises from the chemomechanical interplay of the nonlinear \enquote{local reactor} and Marangoni convection. Ultimately, this gives rise to various forms of self-propelled droplets. To this end, temporarily we only consider the (nondimensional) local reactor given by
\begin{equation}
\begin{aligned}
\dot{\Gamma}_1&=\mathcal{R}+\mathcal{B}_1 = r\left[\delta\Gamma_2\Gamma_1^2-\left(\delta\Gamma_1\right)^3\right]+\beta_1\left[e^{\mu_1}-\delta\Gamma_1\right],\\
\dot{\Gamma}_2&=-\mathcal{R}+\mathcal{B}_2 = -r\left[\delta\Gamma_2\Gamma_1^2-\left(\delta\Gamma_1\right)^3\right]+\beta_2\left[e^{\mu_2}-\delta^{-1}\Gamma_2\right], 
\end{aligned}
\label{eq:local_reactor}
\end{equation}
where the dot denotes the derivative with respect to time and $\delta=a_1/a_2$. To compute the steady states of the system given by (\ref{eq:local_reactor}), we additionally assume that $\beta_2=\mathcal{O}(\varepsilon)$ and $e^{\mu_2}=\mathcal{O}(\varepsilon^{-1})$ with $\varepsilon\ll 1$ and that all other quantities are $\mathcal{O}(1)$. This reflects a slow exchange of $\Gamma_2$ with the bath and a strong energetic bias towards the adsorption of $\Gamma_2$ onto the droplet. These assumptions also capture the parameter choice of Fig.~\ref{fig:selfpropelled_droplets_overview}. Then, the local reactor (\ref{eq:local_reactor}) has a single fixed point which can be determined to order $\mathcal{O}(1)$ as
\begin{equation}
\begin{aligned}
\Gamma_{1,ss}&=\frac{1}{\delta}\left[e^{\mu_1}+\frac{\beta_2}{\beta_1} e^{\mu_2}\right],\\
\Gamma_{2,ss}&=\delta^2\Gamma_{1,ss}+\frac{\beta_2}{r\delta\Gamma_{1,ss}^2} e^{\mu_2},
\end{aligned}
\label{eq:reactor_fp}
\end{equation}
where $\Gamma_{1,ss}, \Gamma_{2,ss}$ are the steady state densities. From Eqs.~(\ref{eq:reactor_fp}) we see that the autocatalyst concentration $\Gamma_{1,ss}$ linearly depends on the chemical driving currents $\beta_1e^{\mu_1},\beta_2 e^{\mu_2}$ whereas the reactant concentration $\Gamma_{2,ss}$ is generally nonlinear in the nonequilibrium forcing [Fig.~\subref{fig:ode_summary}{(e)}]. Notably, this implies due to Eq.~(\ref{eq:linear_Marangoni}) that the nondimensional steady state surface tension $\gamma_{ss}=1-\frac{k_b T}{a_1a_2\gamma_0}(\Gamma_{1,ss}+\Gamma_{2,ss})$ also depends nonlinearly on the driving currents and may increase or decrease when the influx from the bath is increased [Fig.~\subref{fig:ode_summary}{(f)}]. We now consider $\mu_2$ as the main driving force and, for convenience, choose $\beta_2e^{\mu_2}$ as our control parameter (by varying $\mu_2$ and leaving $\beta_2$ fixed). The fixed point (\ref{eq:reactor_fp}) is then rendered unstable in either of two supercritical Hopf bifurcations which are given by the condition (Appendix \ref{app:local_reactor}) 
\begin{equation}
\frac{\partial\left(\Gamma_{1,ss}+\Gamma_{2,ss}\right)}{\partial (\beta_2e^{\mu_2})}=0, \label{eq:reactor_hopf_bifs}
\end{equation}
where both $\Gamma_{1,ss}$ and $\Gamma_{2,ss}$ are understood as functions of $\beta_2e^{\mu_2}$ and of all other parameters (except $\beta_2$ and $\mu_2$) appearing in Eqs.~(\ref{eq:reactor_fp}). That is, the Hopf bifurcations occur at the local extrema of $\gamma_{ss}\left(\beta_2e^{\mu_2}\right)$ [see Fig.~\subref{fig:ode_summary}{(e)}].\footnote{This condition can be understood intuitively by considering a two-component dynamical system with an N-shaped (e.g., cubic) and a vertical nullcline. This is the approximate shape of the nullclines of Eqs.~(\ref{eq:local_reactor}) near the fixed point. In this case, the Hopf bifurcations generically occur when the fixed point crosses the extrema of the N-shaped nullcline. Approximately at these points, also $\Gamma_1+\Gamma_2$ exhibits a local extremum along the branch of steady states that is obtained by shifting the two nullclines with respect to each other.} In between, chemical oscillations are observed [Figs.~\subref{fig:ode_summary}{(b)} and \subref{fig:ode_summary}{(d)}]. Outside of this parameter region, all phase space trajectories converge to the fixed point [Fig.~\subref{fig:ode_summary}{(a)}, Fig.~\subref{fig:ode_summary}{(c)}]. We find that the corresponding reaction-diffusion system shows similar dynamics when $D_1=D_2$ and relaxes to the homogeneous steady state if it is stable. Note that the Hopf bifurcations given by condition (\ref{eq:reactor_hopf_bifs}) correspond exactly to the Hopf bifurcations of the flat film (Fig.~\subref{fig:selfpropelled_droplets_overview}{(c)} shows the first). Therefore, due to our restriction to the purely spinodal case shown in Fig.~\subref{fig:selfpropelled_droplets_overview}{(c)}, we are operating within this stable regime of the local reactor also for the full model (\ref{eq:model}).
\begin{figure}
%\centering
\includegraphics[scale=0.92]{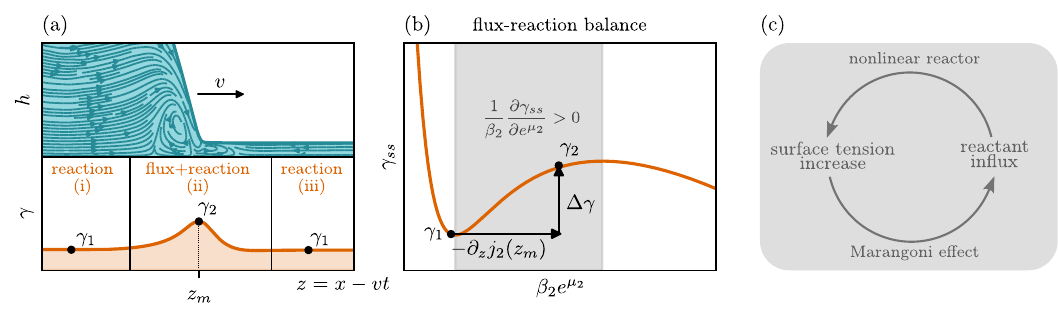}
\caption{(a) Magnification of the front of a self-propelled droplet that moves with velocity $v$. The top panel shows the film thickness $h$ and streamlines represent the velocity field of the liquid in the laboratory frame. The bottom panel shows the surface tension profile $\gamma$. The three regions correspond to (i) the bulk of the droplet, (ii) the advancing contact line and (iii) the adsorption layer. In (i) and (iii), the surface tension $\gamma_1$ is given by the steady state of the local reactor,~i.e., by evaluating $\gamma_{ss}$ at $\beta_2e^{\mu_2}$. At the surface tension maximum $z=z_m$ in (ii), the surface tension $\gamma_2$ is determined as a balance between transport fluxes and chemical reactions and can be obtained by evaluating $\gamma_{ss}$ at $\beta_2e^{\mu_2}\!-\!\partial_z j_2(z_m)$. This is illustrated in (b), where $\gamma_{ss}$ is shown as a function of $\beta_2e^{\mu_2}$. The gray region marks the section between the extrema of $\gamma_{ss}$ (orange dots) where the steady state surface tension increases with increasing influx $-\partial_zj_2$. Note that $\gamma_1$ lies outside of this region. This capacity of the local reactor to increase the surface tension in response to an influx of reactant suggests a general positive feedback loop that is shown in (c).}
\label{fig:flux_reaction_balance}
\end{figure}

We now turn back to the complete spatially extended system and consider a self-propelled droplet in the comoving frame $z=x-vt$. If the droplet moves with constant velocity $v$, the surfactant profiles are given by 
\begin{equation}
\begin{aligned}
0&=-\partial_z j_1+\mathcal{R}+\mathcal{B}_1-v\partial_z\Gamma_1, \\
0&=-\partial_z j_2-\mathcal{R}+\mathcal{B}_2-v\partial_z\Gamma_2.
\end{aligned}
\label{eq:comoving_frame}
\end{equation}
We consider self-propelled states given by Eqs.~(\ref{eq:comoving_frame}) in three different regions (i) to (iii) shown in Fig.~\subref{fig:flux_reaction_balance}{(a)}. Region (i) is located away from the contact line region and corresponds to the bulk of the droplet. The flow is nearly laminar and matter is \enquote{passed through} to the contact line region. Consequently, no surfactant accumulates due to transport and we have $\partial_z j_{1}=\partial_z j_{2}=0$. We find that the surfactant profile is uniform in this region such that $v\partial_z\Gamma_{1}=v\partial_z\Gamma_{2}=0$. The concentrations are then given by the steady state of the local reactor (\ref{eq:reactor_fp}) and we denote the respective surface tension by $\gamma_1$. Region (iii) comprises the adsorption layer far away from the contact line region. Here, the film thickness is constant ($h=1$, or $h=h_a$ in dimensional units) and the surfactant coverage is uniform.  As before, we then have $\partial_zj_{i}=v\partial_z\Gamma_{i}=0$ with $i=1,2$ and the local concentrations are again given by Eqs.~(\ref{eq:reactor_fp}) with the surface tension $\gamma_1$. Finally, we turn to the contact line region (ii). There, the liquid flow is dominated by two vortices that are \enquote{squeezed} into the contact line region due to the strong left-right symmetry breaking. The weaker vortex reaches into the adsorption layer [Fig.~\subref{fig:flux_reaction_balance}{(a)}]. Here, the fluxes $j_{1}$ and $j_2$ are generally not constant and there exist strong surface tension gradients. However, we only consider the local maximum of the surface tension profile at $z=z_m$ with the surface tension $\gamma_2$ and where $\partial_z\gamma=0$ and \textit{approximately} $\partial_z\Gamma_1=\partial_z\Gamma_2=0$ (since the local extrema of $\Gamma_1$ and $\Gamma_2$ do not coincide exactly). In particular when the droplet speed is not too large, the contributions $-v\partial_z\Gamma_1, -v\partial_z\Gamma_2$ are then negligible. We additionally observe in time simulations that the transport contributions $-\partial_z j_1$ are small near the surface tension peak. This can be explained by observing that $\Gamma_1$ is enriched in the contact line region [see Fig.~\subref{fig:selfpropelled_droplets_overview}{(a)}] such that diffusion of $\Gamma_1$ opposes the advective fluxes (which transport surfactant \textit{into} the contact line region). Diffusive and advective contributions then effectively cancel near the peak. Note that the difference in coverage between $\Gamma_1$ and $\Gamma_2$ (particularly near the advancing contact line) originates from differences in $\beta_1, \beta_2$ and $\mu_1,\mu_2$ as well as from the distinct roles of the two species in the autocatalysis (\ref{eq:reaction_autocat}) since we assume $D_1=D_2$ and $\delta=1$. We further discuss these approximations in Appendix~\ref{app:approximations_flux_reaction_balance}. As a result, at the local maximum of the surface tension profile at $z=z_m$ we have the balance equations
\begin{equation}
\begin{aligned}
0&=\mathcal{R}+\mathcal{B}_1,\\
0&=-\mathcal{R}+\mathcal{B}_2-\partial_z j_2.
\end{aligned}
\label{eq:comoving_frame_gamma_max}
\end{equation}
We recognize that Eqs.~(\ref{eq:comoving_frame_gamma_max}) correspond to the steady state equations of the local reactor (\ref{eq:local_reactor}), augmented by the spatial transport of reactant $\Gamma_2$. We are now interested in steady states with $-\partial_z j_2(z_m)>0$ at the peak while $\Delta\gamma = \gamma_2-\gamma_1>0$. In this scenario matter is continuously advected into the contact line region due to a simultaneously maintained gradient in surface tension between the three discussed regions, which must hold for self-propelled droplets. Using Eqs.~(\ref{eq:comoving_frame_gamma_max}), we see that $-\partial_z j_2(z_m)$ (which can be directly determined from time simulations) acts simply as an additional driving term that may be added to the control parameter,~i.e., at the surface tension peak the surface tension $\gamma_2$ can be found by evaluating $\gamma_{ss}$ as a function of $\beta_2e^{\mu_2}\!-\!\partial_z j_2$. Because $\gamma_{ss}$ changes $\textit{non-monotonically}$ with the control parameter, there is a region of the curve where $\gamma_{ss}$  increases with the driving current,~i.e., the total mass in the reactor \textit{decreases} with an increasing influx of reactant [Fig.~\subref{fig:flux_reaction_balance}{(b)}]. We then find that $\Delta\gamma=\gamma_{ss}\left(\beta_2 e^{\mu_2}-\partial_z j_2(z_m)\right)-\gamma_{ss}\left(\beta_2 e^{\mu_2}\right)>0$ may coincide with $-\partial_z j_2(z_m)>0$ for sufficiently large transport contributions (note that the stability of the local reactor does \textit{not} reflect the stability of the self-propelled state). Physically, this effect can be summarized as follows. The local reactor may overcompensate additional influxes of the reactant $\Gamma_2$ from neighboring regions on the free surface by a net removal of surfactant from the droplet, thereby maintaining a locally increased surface tension. In consequence, matter is continuously advected into the advancing contact line region and the droplet moves across the substrate. Self-propelled droplets as shown in Fig.~\ref{fig:selfpropelled_droplets_overview} and Fig.~\ref{fig:flux_reaction_balance} therefore represent a balance between chemical reactions and transport fluxes that is sustained due to an interplay of the nonlinear local reactor and the Marangoni effect.

This interaction between the Marangoni effect and the local reactor relies on the capacity of the local reactor to increase the surface tension in response to a reactant influx. More generally, this suggests a positive feedback loop where locally induced Marangoni flows cause increases in surface tension via the local reactor that in turn enhance these flows [Fig.~\subref{fig:flux_reaction_balance}{(c)}]. Surface tension gradients can therefore be quickly excited,~e.g., by perturbations of the local flow. We can then expect more complex forms of self-propulsion when surface tension gradients are excited away from the advancing contact line, which may cause droplets to stop or even reverse their direction of propagation. We investigate such states in the following and study the bifurcations that give rise to various forms of droplet motility.

\section{Partial bifurcation study and complex forms of drop motility}\label{sec:bifurcation_analysis}

\begin{figure}
%\centering
\includegraphics[scale=0.95]{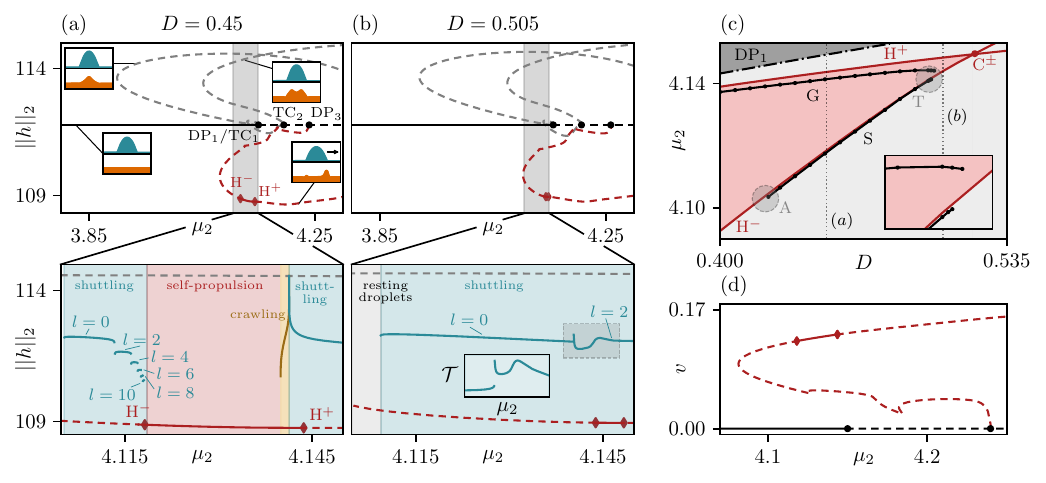}
\caption{(a) Partial bifurcation diagram of resting and moving droplets as a function of the chemical potential $\mu_2$ at $D=0.45$ with the L$_2$-norm $\vert\vert h \vert\vert_2\!=\!\left(\int h^2 \text{d}x\right)^{1/2}$ as solution measure (period-average of $\vert\vert h \vert\vert_2$ for time-periodic states). Linearly stable states [unstable states] are represented as solid lines [dashed lines]. Only the most relevant bifurcations and branches are shown. At small $\mu_2$ the resting-drop base state (black line) is linearly stable. It is then rendered unstable in a series of bifurcations DP$_1$, TC$_1$ (cf. magnification in Fig.~\ref{fig:bifurcation_diagram_zoom}). The simple traveling droplets bifurcate at a drift-pitchfork bifurcation DP$_3$. The emerging states (red line) are unstable, but a linearly stable section is limited by two Hopf bifurcations H$^-$, H$^+$ (red diamonds). The insets schematically show exemplar droplet and surface tension profiles. The gray region is magnified in the bottom panel where colored regions illustrate what kind of drop motion is obtained in time simulations initialized with a flat film. (b) Analogous bifurcation diagram for $D\!=0.505$. In the magnified region only shuttling is observed in time simulations. The small inset in the bottom panel shows the temporal period $\mathcal{T}$ for both branches in the marked region. (c) Two-parameter bifurcation diagram in the $(\mu_2,D)$-plane with $D\!=D_1\!=\!D_2$. Red lines mark the Hopf bifurcations H$^-$, H$^+$ that cross at C$^\pm$. In the red region, the traveling states are linearly stable. The lines G and S correspond to the gluing bifurcation and the Shilnikov bifurcation. Black points mark data obtained from time simulations. The regions T and A represent possible termination points. The inset shows a magnification near the suspected T-point. Above DP$_1$ (dash-dotted line) the base state is unstable. The two vertical dotted lines indicate the parameters of the bifurcation diagrams shown in (a) and (b). (d) Velocity of the traveling states in (a). The remaining parameters for all panels are $\mu_1=-1.4, r=0.3, \beta_1=2, \beta_2=0.01, \delta=1, W=10$ with a mean film thickness of $\bar{h}=7$. The computational domain is $[0,100]$ with periodic boundaries.}
\label{fig:bifurcation_diagrams}
\end{figure}

\begin{figure}
%\centering
\includegraphics[scale=1]{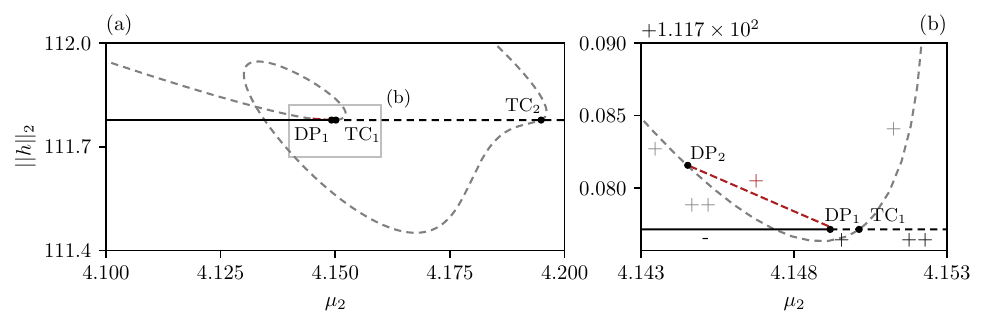}
\caption{(a) Magnification of the bifurcation structure of Fig.~\subref{fig:bifurcation_diagrams}{(a)} near the first instabilities of the base branch,~i.e., near DP$_1$, TC$_1$ and TC$_2$. Panel (b) shows a further magnification of the region marked in (a). The base state is first rendered unstable via the subcritical drift-pitchfork bifurcation DP$_1$. The emerging branch of traveling states (red line) connects at a supercritical drift-pitchfork DP$_2$ to the branch that bifurcates backwards at the transcritical bifurcation TC$_1$. The number of unstable eigenvalues (EV) for each branch is given ($-$: linearly stable, $+$: one unstable EV, $++$: two unstable EVs).}
\label{fig:bifurcation_diagram_zoom}
\end{figure}
\subsection{Onset of self-propulsion}\label{sec:onset_selfpropulsion}
To explore the bifurcations that ultimately result in complex types of drop motility, we employ numerical continuation \citep{KrauskopfOsingaGalan-Vioque2007, DWCD2014ccp,EGUW2019springer} using the package pde2path \cite{Uecker2021} (see Appendix \ref{app:numerical_methods}). We first investigate bifurcations from the base branch of resting droplets that are uniformly covered by surfactants. At $\mu_1\!=\!\mu_2$, this state corresponds to thermodynamic equilibrium. We additionally assume equal diffusion of both surfactants with $D_1\!=\!D_2\!=\!D$. A typical bifurcation diagram with control parameter $\mu_2$ at fixed $\mu_1\neq 0$ is presented in Fig.~\subref{fig:bifurcation_diagrams}{(a)}. Note that we restrict ourselves to states that are relevant to what is observed in time simulations and therefore only show a small selection of all existing branches and bifurcations. As $\mu_2$ is increased, the branch of uniformly covered drops is rendered unstable in a combination of a drift-pitchfork bifurcation DP$_1$ and a transcritical bifurcation TC$_1$ that occur in immediate succession.\footnote{Note that in Ref.~\citep{VoTh2024jem} this structure was erroneously identified as a higher-multiplicity pitchfork bifurcation.} The fully resolved bifurcation structure near DP$_1$ and TC$_1$ is shown in Fig.~\ref{fig:bifurcation_diagram_zoom}. First, a branch of parity-broken unstable traveling states emerges subcritically at DP$_1$. This branch then connects supercritically at a drift-pitchfork DP$_2$ to an unstable branch of resting symmetric states that bifurcates from the base branch at TC$_1$ and connects to it again at another transcritical bifurcation TC$_2$ at a larger value of $\mu_2$. This series of bifurcations comprising DP$_1$, TC$_1$, DP$_2$ and TC$_2$ essentially produces two branches of unstable symmetric resting states that continue to exist far away from the bifurcations. These states feature either one or two surface-tension peaks near the droplet center [Fig.~\subref{fig:bifurcation_diagrams}{(a)}]. At larger driving, a branch of traveling states emerges at a subcritical drift-pitchfork DP$_3$ with a linearly stable section that is limited by two subcritical Hopf bifurcations H$^-$ and H$^+$ [Figs.~\subref{fig:bifurcation_diagrams}{(a)} and \subref{fig:bifurcation_diagrams}{(d)}]. Note that H$^-$ and H$^+$ are unrelated to the Hopf bifurcations of the local reactor given by condition (\ref{eq:reactor_hopf_bifs}). This section of linearly stable states corresponds to the simple self-propelled droplets described above in Sec.~\ref{sec:simple_traveling}.\footnote{In Fig.~\subref{fig:bifurcation_diagrams}{(a)} this linearly stable section is comparatively small. However, it can be broadened significantly (up to an order of magnitude) by decreasing the diffusion constant $D$ as indicated in Fig.~\subref{fig:bifurcation_diagrams}{(c)}.} The corresponding branch features several saddle-node bifurcations which are, however, only visible as \enquote{kinks} in Figs.~\subref{fig:bifurcation_diagrams}{(b)} and \subref{fig:bifurcation_diagrams}{(d)}. We remark that the drift-pitchfork DP$_3$ occurs beyond a Hopf bifurcation of the base branch (not shown) that corresponds to the first Hopf instability of the flat film [Fig.~\subref{fig:selfpropelled_droplets_overview}{(c)}] (i.e., the first Hopf instability of the local reactor). Nevertheless, the linearly stable section of the branch of self-propelled droplets lies fully within the parameter region where the flat film is only spinodally unstable.

\subsection{Crawling and shuttling droplets}\label{sec:crawling_shuttling}
\begin{figure}
%\centering
\includegraphics[scale=0.96]{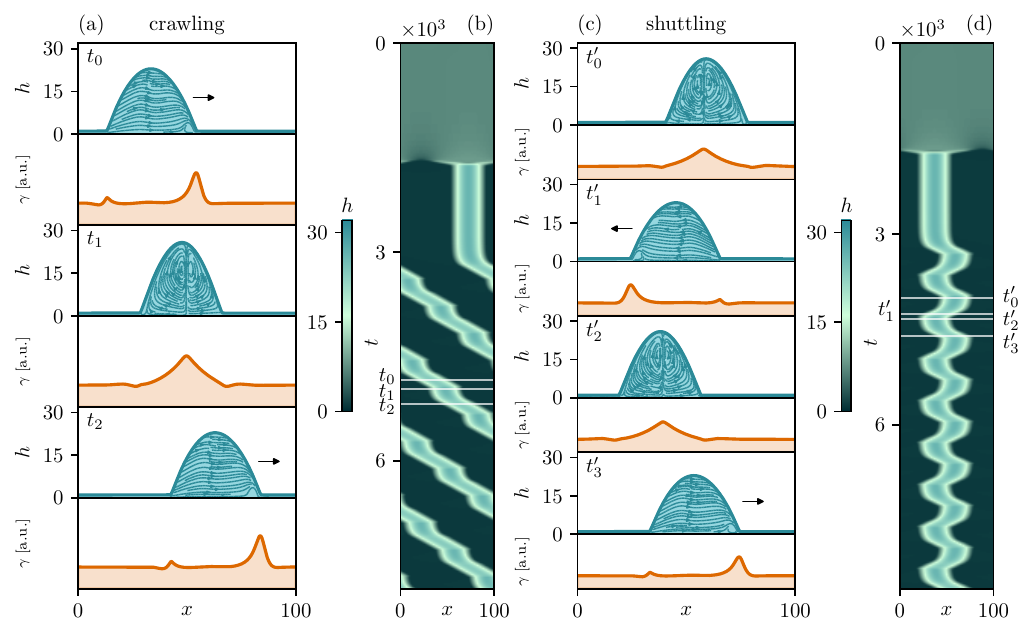}
\caption{Comparison of crawling and shuttling droplets. Panels (a) and (b) respectively show snapshots and a space-time plot for a right-crawling droplet. At times $t_0, t_2$ the droplet moves as indicated by the arrows and at time $t_1$ the droplet rests. The streamlines represent the velocity field of the bulk liquid in the laboratory frame. After an initial dewetting phase, a single droplet spontaneously breaks left-right symmetry and begins to crawl across the substrate. The white horizontal lines in (b) denote the times presented in (a). Panels (c) and (d) show a shuttling droplet which reverses its direction of propagation at times $t'_0$ and $t'_2$ and travels at times $t'_1, t'_3$. The parameters are $\mu_1=-1.4, r=0.3, \beta_1=2, \beta_2=0.01, \delta=1, D_1=D_2=0.45, W=10$ with a mean film thickness of $\bar{h}=7$. For (a), (b) $\mu_2=4.14$ and for (c), (d) $\mu_2=4.145$. The computational domain is $[0,100]$ with periodic boundaries. See also Supplemental Videos 2 and 3.}
\label{fig:crawling_shuttling_overview}
\end{figure}

We next focus on the parameter region near the primary instability of the base branch where also the simple traveling states occur [gray area in Fig.~\subref{fig:bifurcation_diagrams}{(a)}]. In this region, we further observe strikingly complex forms of self-propulsion that we study using direct numerical simulations. First, we consider the parameter region near the destabilizing Hopf bifurcation H$^+$. We typically find two forms of droplet motion (Fig.~\ref{fig:crawling_shuttling_overview}). On the one hand, droplets \enquote{crawl} across the substrate by periodically forming a single surface tension peak near the droplet center that is subsequently advected to the advancing contact line region [Figs.~\subref{fig:crawling_shuttling_overview}{(a)} and \subref{fig:crawling_shuttling_overview}{(b)}, Supplemental Video 2]. This results in phases of uniform motion interrupted by abrupt stops. On the other hand, droplets \enquote{shuttle} between two points on the homogeneous substrate by reversing their direction of travel after stopping [Figs.~\subref{fig:crawling_shuttling_overview}{(c)} and \subref{fig:crawling_shuttling_overview}{(d)}, Supplemental Video 3]. For both types of motion, we find that the fluid flow at times long after a stop (times $t_0, t_2,t'_1, t'_3$ in Fig.~\ref{fig:crawling_shuttling_overview}) is nearly identical to the simple traveling states shown in Fig.~\ref{fig:selfpropelled_droplets_overview}. Perturbations of this flow then lead to an explosive generation of a new surface tension peak near the droplet center due to the positive feedback loop outlined in Sec.~\ref{sec:simple_traveling}. Whether the droplet reverses its direction is determined by the exact position of the newly generated surface tension peak. When it is created slightly on the side of the droplet center that is opposite to the already existing peak the drop motion changes direction, otherwise it retains its direction of travel. We also find that during the advection of the new surface tension peak to the advancing contact line region, Eqs.~(\ref{eq:comoving_frame_gamma_max}) are still valid when $-\partial_z j_1$ is taken into account (Supplemental Video 4). This implies that during this phase, the local reactive dynamics is fast as compared to hydrodynamic transport.

\begin{figure}
%\centering
\includegraphics[scale=0.93]{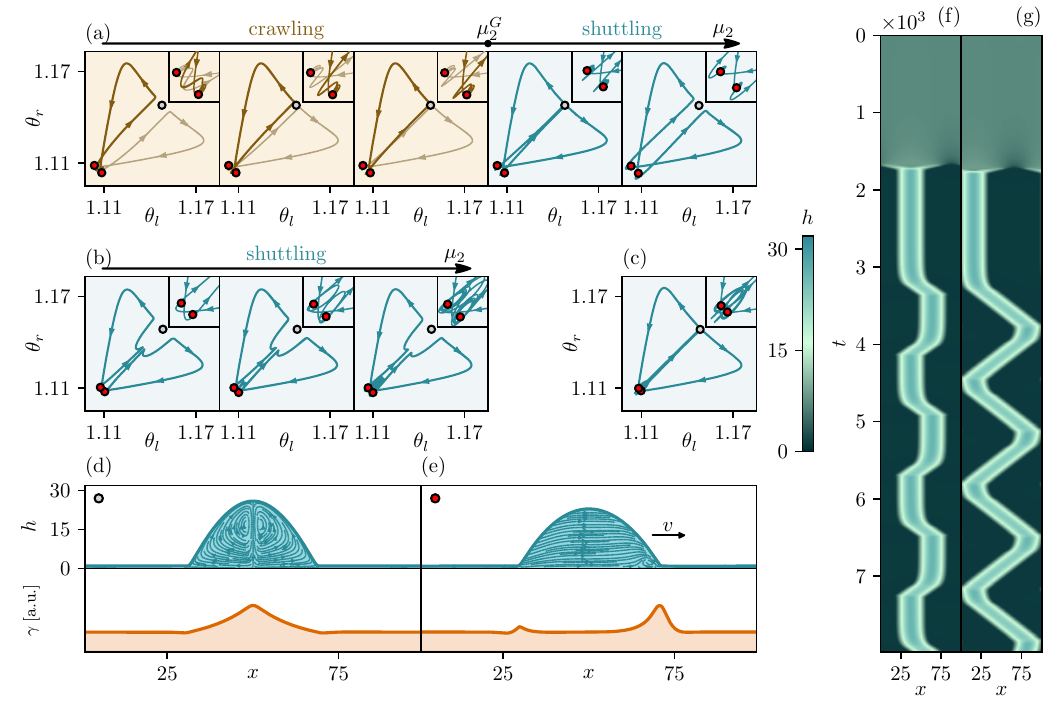}
\caption{\setstretch{1.0}(a) Representation of crawling and shuttling droplets as periodic orbits in the plane spanned by the right and left contact angles $\theta_r$ and $\theta_l$. The two simple traveling states and the one-peak resting state are marked as red points and gray points, respectively. The arrows represent motion along the orbit and insets zoom into the region near the traveling states. For crawling droplets, right-crawling (strong lines) and left-crawling droplets (weak lines) are shown. They are related by reflection. With increasing $\mu_2$, the crawling orbits move closer to the resting state. At a critical driving $\mu_2^G$ they form a \enquote{two-winged} shuttling orbit via a gluing bifurcation. The shuttling orbits closely pass both traveling states. The driving forces are $\mu_2=4.140, 4.1405, 4.141, 4.1416,4.147$ and $\mu_2^G\approx 4.1413$. The diffusion constants are $D_1=D_2=0.45$. (b) Series of shuttling orbits that successively \enquote{wind around} the traveling states. The driving forces are $\mu_2=4.112, 4.115, 4.1181$ with $D_1=D_2=0.45$. (c) Shuttling orbit at $\mu_2=4.14325$ and $D_1=D_2=0.503$ that closely passes by the resting state and also winds around both traveling states. (d) and (e) show height and surface tension profiles of the symmetric resting and the traveling states shown as fixed points in (a)-(c). The symmetric state is always unstable. The streamlines represent the velocity of the bulk liquid in the laboratory frame. (f), (g) Space-time representations of shuttling droplets that (f) are close to the gluing bifurcation and therefore rest longer and (g) form many loops around the traveling states and thus travel for long phases. The driving forces are (f) $\mu_2=4.14136$ and (g) $\mu_2=4.1183$ with $D_1=D_2=0.45$. The remaining parameters for all panels are as in Fig.~\ref{fig:crawling_shuttling_overview}. See also Supplemental Videos 5 and 6.}
\label{fig:contact_angle_trajectories}
\end{figure}

Both crawling and shuttling motion can be understood as a periodic transition between an unstable resting and an unstable moving conformation [Figs.~\subref{fig:contact_angle_trajectories}{(d)} and \subref{fig:contact_angle_trajectories}{(e)}]. The moving conformation of the drop closely resembles the (unstable) simple traveling states.\footnote{Note that there exist parameter regions, notably near H$^+$, where either crawling or shuttling and the simple traveling state are multistable [Fig.~\subref{fig:bifurcation_diagrams}{(a)}]. In this case, it is the unstable (quasi)-time periodic state that emerges subcritically at the nearby Hopf bifurcation H$^+$ that prevents the dynamics from converging to the simple traveling state.} The resting conformation corresponds to the symmetric resting droplet with one surface tension peak. This state is always unstable [Fig.~\subref{fig:bifurcation_diagrams}{(a)}] and any arbitrarily small breaking of the left-right symmetry induces droplet motion. Crawling and shuttling motion can both be conveniently represented as periodic orbits in a reduced phase space where the orbits closely pass the resting and the moving states. Therefore, we refer to both forms of motion as periodic (the period corresponds to the time of a single orbit roundtrip). We use a projection onto the ($\theta_l, \theta_r$)-plane, where $\theta_l$ and $\theta_r$ are the left and right contact angles, respectively. They are determined from the slopes at the inflection points of $h$ (Appendix \ref{app:numerical_methods}). Examples are presented in Fig.~\subref{fig:contact_angle_trajectories}{(a)}. For crawling droplets, the orbits may be divided into three phases that represent different stages of the motion. First, the trajectory passes by either the right- ($\theta_l>\theta_r$) or the left-traveling droplet state ($\theta_r>\theta_l$). This corresponds to a right- or left-crawling droplet, respectively. Second, the trajectory is quickly expelled from the region near the traveling state and approaches the symmetric resting state ($\theta_l=\theta_r$). In this phase, a new surface tension peak is generated near the droplet center, the droplet abruptly stops and the dynamics slows down. As the new surface tension peak appears near the droplet center, the orbit also crosses the diagonal $\theta_l=\theta_r$,~i.e., the advancing contact angle becomes greater than the receding one as the trajectory approaches the resting state. This does not correspond to a direction reversal but reflects a change of the droplet shape as the new surface tension peak appears. Third, as the newly generated peak is advected to the advancing contact line the trajectory departs from the unstable resting state and again approaches the moving state. 
\begin{figure}
%\centering
\includegraphics[scale=1]{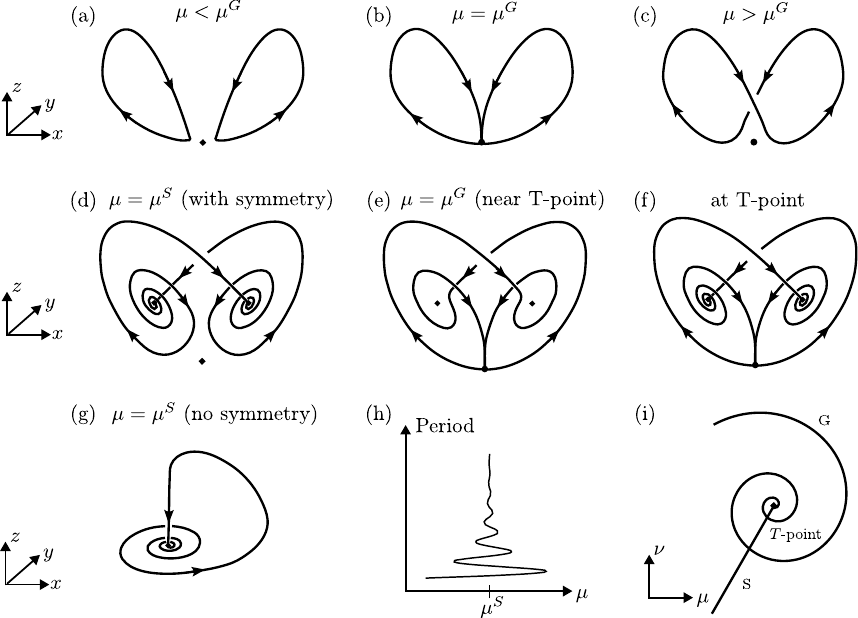}
\caption{Schematic representation of phase space behavior related to the transitions between crawling and shuttling states. (a)-(c) Gluing bifurcation in a three-dimensional dynamical system with a symmetry with the control parameter $\mu$. Two initially separate periodic orbits in (a) form a double homoclinic loop at the central fixed point in (b) and finally merge into a single symmetric orbit shown in (c). (d) Symmetric heteroclinic connections between two non-central fixed points at a Shilnikov bifurcation. (e) Same as (b) near a T-point. The homoclinic connections closely pass by the non-central fixed points. (f) Symmetric heteroclinic loops between the central and the non-central fixed points at a T-point. (g) Homoclinic loop at a single fixed point at a Shilnikov bifurcation. (h) Schematic bifurcation diagram of a periodic orbit that approaches a homoclinic connection [pair of heteroclinic connections] at $\mu_S$ when the Shilnikov condition is fulfilled (see main text). The corresponding branch snakes around $\mu_S$ and the period tends to infinity as $\mu\rightarrow\mu_S$. (i) Sketch of a two-parameter bifurcation diagram in the $(\mu,\nu$)-plane near the T-point shown in (f). The loci of the double homoclinic loop [see (e), gluing bifurcation G] spiral into the T-point while the loci of the pair of heteroclinic orbits [see (d), Shilnikov bifurcation S] approach it in a straight line.}
\label{fig:bifurcation_sketches}
\end{figure}

\begin{figure}
%\centering
\includegraphics[scale=1]{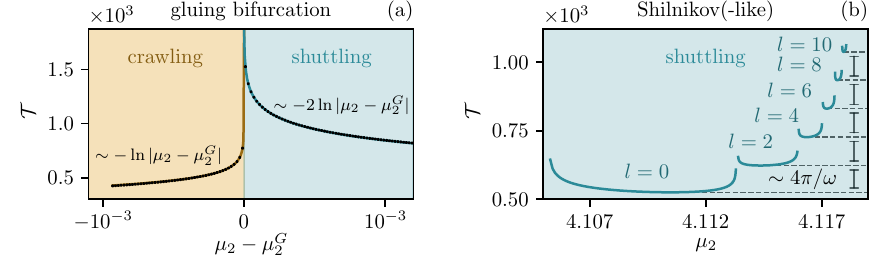}
\caption{(a) Scaling of the temporal period $\mathcal{T}$ of crawling and shuttling motion as the gluing bifurcation is approached. The period corresponds to a single orbit roundtrip in the ($\theta_l,\theta_r$)-plane. Black points denote points measured in time simulations. The solid lines show logarithmic fits with $\mathcal{T}=-A\ln\vert\mu_2-\mu_2^G\vert+B$ where $A, B$ and $\mu_2^G$ are fit parameters. In particular, $\mu_2^G\approx 4.14133$. (b) Periods of the first few pairs of loops that are formed in a Shilnikov(-like) mechanism for shuttling droplets. Only the stable parts of the branches are shown. From one branch to the next, $\mathcal{T}$ differs by approximately $4\pi/\omega$. The total number of loops is denoted by $l$. The remaining parameters are as in Fig.~\ref{fig:crawling_shuttling_overview}.}
\label{fig:period_scalings}
\end{figure}

\begin{figure}
%\centering
\includegraphics[scale=1]{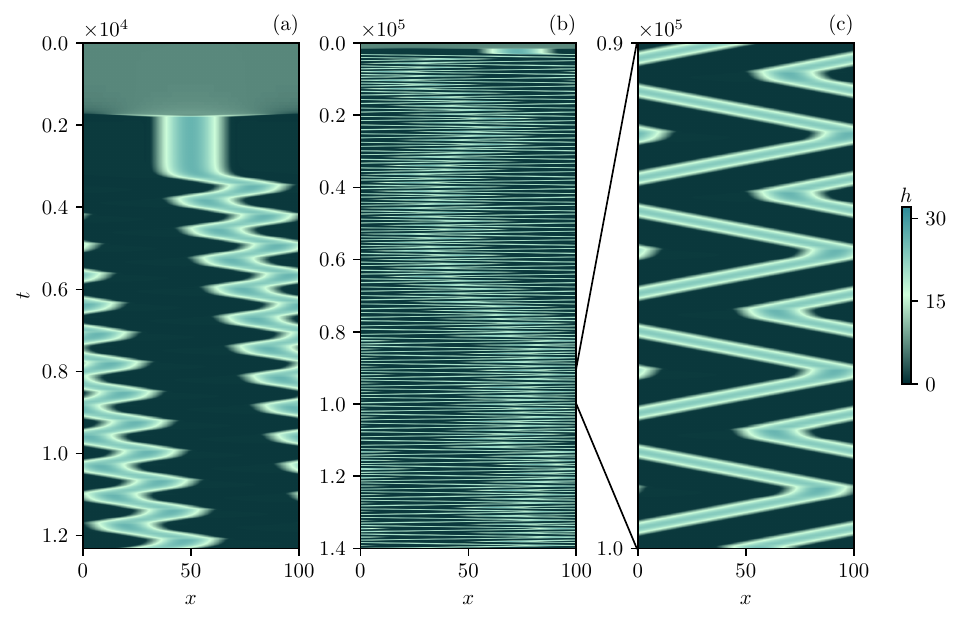}
\caption{Space-time plots for more complex variants of shuttling motion. Panel (a) shows asymmetric shuttling motion with different numbers of loops around the left- and right-traveling state. Panel (b) shows irregular shuttling motion over a long time span. The droplet aperiodically reverses its direction and each section of the motion corresponds to a different number of loops around one of the traveling states. Panel (c) shows a magnification of the marked section in (b) thereby illustrating the small changes in the loci of direction reversal. The chemical potentials are (a) $\mu_2=4.11599$ and (b) $\mu_2=4.11854$. The remaining parameters are as in Fig.~\ref{fig:crawling_shuttling_overview}. See also Supplemental Videos 7 and 8.}
\label{fig:exotic_spacetime_plots}
\end{figure}

When the chemical potential $\mu_2$ is decreased, the stable section of the branch of crawling states ends in what we believe to correspond to a saddle-node bifurcation [Fig.~\subref{fig:bifurcation_diagrams}{(a)}]. For increased values of $\mu_2$, trajectories corresponding to crawling motion come gradually closer to the one-peak resting state,~i.e., droplets rest longer before resuming motion. At a critical driving $\mu_2^G$ the orbits of left- and right-crawling droplets meet at this state and beyond this value only shuttling droplets exist. They then represent a single orbit that passes by both the left- and the right-traveling droplet state [Fig.~\subref{fig:contact_angle_trajectories}{(a)}]. Close to this transition we find that the one-peak resting droplet only has one unstable eigenvalue and that the leading eigenvalues (closest to the imaginary axis) are real,~i.e., the resting droplet represents a saddle point in phase space. This transition is a gluing bifurcation \citep{GaGT1988n}, where a pair of periodic orbits (here related by reflection symmetry $x\rightarrow -x$) forms a double homoclinic loop at a saddle point and merge afterwards to form a single orbit. The transition is schematically shown in Figs.~\subref{fig:bifurcation_sketches}{(a)} to \subref{fig:bifurcation_sketches}{(c)}. The period then diverges logarithmically as the critical parameter value is approached \cite{PaPe2001pre}. We numerically confirm this scaling for the transition from crawling to shuttling droplets in Fig.~\subref{fig:period_scalings}{(a)}. In other contexts, gluing bifurcations have been reported,~e.g., for models of optothermal cavities \cite{HFPP1998pre} and periodically forced Taylor-Couette flows \cite{LoMa2000prl}, and play crucial roles in some transitions to chaos \cite{DeKr1999prl,KuKo1982pla,AbPM2001prl}. For highly simplified models in the form of ordinary differential equations,~e.g., for reaction-diffusion fronts \cite{PaPe2001pre} and active deformable particles \cite{TaOh2016el}, they have been shown to give rise to direction reversing motion as also discussed here. 

Shuttling states are also found near the stabilizing Hopf bifurcation H$^-$ [Fig.~\subref{fig:bifurcation_diagrams}{(a)}]. However, they display strikingly different behavior when the driving is increased. A typical sequence of trajectories is presented in Fig.~\subref{fig:contact_angle_trajectories}{(b)}. As $\mu_2$ is increased, the orbit \enquote{winds around} both traveling states, successively forming more loops (the total number of loops is denoted by $l$). Droplets then maintain their direction of travel for a longer time as the number of loops increases while both contact angles slightly oscillate. This is in stark contrast to the situation shown in Fig.~\subref{fig:contact_angle_trajectories}{(a)}, where droplets rest longer as the gluing bifurcation is approached. These two scenarios are also compared in Figs.~\subref{fig:contact_angle_trajectories}{(f)} and \subref{fig:contact_angle_trajectories}{(g)} (Supplemental Videos 5 and 6). We remark that particularly in cases like in Fig.~\subref{fig:contact_angle_trajectories}{(g)}, the droplet is not restricted by the domain boundaries,~i.e., it may also turn around \textit{after} crossing the periodic boundary. We find that with each additional pair of loops around the traveling states, the temporal period of a single shuttle (one \enquote{back-and-forth}) increases approximately by $4\pi/\omega$ (each loop contributes $2\pi/\omega$), where $\lambda_{1,2}=\rho\pm i\omega$ with $\rho>0$ are the eigenvalues associated with the Hopf instability H$^-$ (they are the only unstable eigenvalues). The leading stable eigenvalue $\lambda_3<0$ is real with $\vert\rho/\lambda_3\vert<1$.\footnote{This is a direct consequence of the nearby Hopf instability H$^-$, where $\rho=0$.} Note that $\omega$ remains approximately constant over the parameter region where we observe shuttling. We show the shuttling period over $\mu_2$ in Fig.~\subref{fig:period_scalings}{(b)} for the first few pairs of loops. We expect that the stable part of each branch is limited by two saddle-node bifurcations as the period does not diverge at the ends of the branches. They may also be limited,~e.g, by period doubling bifurcations on one side as in Ref.~\citep{Ko1995pd}.

We believe that this dynamics is caused by the presence of a pair of heteroclinic connections between the traveling states which exists for some nearby value of $\mu_2$. For the present case with $\vert\rho/\lambda_3\vert < 1$, it is known that a homoclinic loop gives rise to a branch of time-periodic states which snakes around homoclinicity in a succession of infinitely many saddle-node bifurcations that accumulate at homoclinicity and where the period along the branch increases by a constant with each saddle-node \cite{ShilnikovShilnikovTuraevChua2001,GlSp1984jsp} [Figs.~\subref{fig:bifurcation_sketches}{(g)} and \subref{fig:bifurcation_sketches}{(h)}]. This scenario is also known as a Shilnikov bifurcation and is related to various transitions to chaos \cite{ArCT1980pla,MTKW1983n, KMTW1986jfm}. Similar phenomena occur in systems with an additional symmetry (here the reflection symmetry $x\rightarrow -x$) where a pair of heteroclinic connections takes the role of a single homoclinic one \cite{GlSp1984jsp} [Fig.~\subref{fig:bifurcation_sketches}{(d)}]. We observe that in the case of Fig.~\subref{fig:period_scalings}{(b)} there is no multistability between the stable parts of each branch and that each stable section \enquote{folds upwards} at \text{both} ends, in contrast to what is expected for a single branch approaching heteroclinicity [Fig.~\subref{fig:bifurcation_sketches}{(h)}]. However, since standard analyses like the one in Ref.~\cite{GlSp1984jsp} are only strictly valid for parameter values in some unspecified small neighborhood of heteroclinicity, there is no direct contradiction. The stable branch segments shown in Fig.~\subref{fig:period_scalings}{(b)} could then be connected without any hysteresis or form separate isolas.

\subsection{Long-time drift and random motion}
In the very small regions between the stable sections in Fig.~\subref{fig:period_scalings}{(b)} we observe shuttling with unequal numbers of loops around the left- and right-traveling states. This induces a long-time effective drift of the droplet [Fig.~\subref{fig:exotic_spacetime_plots}{(a)}, Supplemental Video 7]. Corresponding asymmetric periodic orbits also emerge in the context of Shilnikov bifurcations with symmetry \cite{GlSp1984jsp}. Further, for driving forces that correspond to a large number of loops ($l>30$) we also find shuttling motion, where the number of loops changes irregularly with time. This results in what appears to be long-time random drop motion [Fig.~\subref{fig:exotic_spacetime_plots}{(b)}, Supplemental Video 8]. This irregular direction-reversal dynamics is likely connected to the presence of the pair of (Shilnikov) heteroclinic connections between the right- and left-traveling states, but in other systems with reflection and translation symmetry it may also result from symmetry-increasing bifurcations of already existing chaotic attractors \citep{KnLM1999pla}. Because the relevant parameter regions for both cases are extremely small, here we do not discuss them further. 

Lastly, we remark that quite similar bifurcation cascades, where the Shilnikov condition holds, have been reported for systems of ordinary differential equations \citep{KoGa1991jpc,KoGa1992jcp,Ko1995pd}. In Refs.~\citep{KoGa1991jpc,KoGa1992jcp} they were also named \enquote{incomplete homoclinic scenarios} due to the apparent absence of a Shilnikov homoclinic orbit. Nevertheless, in particular for the Koper model \citep{Ko1995pd}, the existence of such a homoclinic orbit was recently demonstrated \citep{GuLi2015sjads}.

\subsection{Organization around higher-codimension bifurcations}\label{sec:higher_codim}
Next, we examine how the foregoing bifurcations change when a second parameter, the diffusion constant $D\!=\!D_1\!=\!D_2$, is varied. The loci of the bifurcations discussed above are tracked in the $\left(\mu_2,D\right)$-plane and presented in Fig.~\subref{fig:bifurcation_diagrams}{(c)}. The locus of the primary bifurcation DP$_1$ from the base branch shows that the uniformly covered resting droplet is linearly stable for most of the shown parameter region. The Hopf bifurcations H$^+$ and H$^-$ interchange positions at C$^\pm$, rendering the simple traveling droplets unstable for all values of $\mu_2$. Of particular interest are the loci of the gluing bifurcation G and of the inferred Shilnikov bifurcation S.\footnote{For a given $D$, we determine $\mu^G$ from the divergence of the period at the gluing bifurcation. We approximate $\mu^S$ by the location of the $l\!=\!8$-branch.} When $D$ is increased, eventually the distance of the loci of G and S becomes very small [Fig.~\subref{fig:bifurcation_diagrams}{(c)}, region T]. Beyond this region, the one-parameter bifurcation diagram  at fixed $D$ transforms drastically and only two branches of shuttling states exist with either $l\!=\!0$ or $l\!=\!2$ loops, forming a continuous parameter region in which shuttling is observed [Fig.~\subref{fig:bifurcation_diagrams}{(b)}]. Notably, the period remains finite. It is instructive to consider the representation of states in the ($\theta_l,\theta_r$)-plane in the region T [Fig.~\subref{fig:contact_angle_trajectories}{(c)}]. Typical trajectories,~e.g., near the gluing bifurcation additionally show several loops, indicating that the homoclinic connections at the gluing bifurcation also closely pass by the traveling states [Fig.~\subref{fig:bifurcation_sketches}{(e)}]. We can then expect that there exists a point in the ($\mu_2,D$)-plane where the homoclinic connections collide with the traveling states, or equivalently, where the heteroclinic connections at the Shilnikov bifurcation meet at the central one-peak resting state. Then, the central resting state is simultaneously connected to both traveling states in a heteroclinic loop [Fig.~\subref{fig:bifurcation_sketches}{(f)}]. The corresponding codimension-2 point where G and S terminate is called a T-point \citep{GlSp1986jsp,Byko1993pd}. The two-parameter bifurcation diagram near this point is expected to be similar to Fig.~\subref{fig:bifurcation_sketches}{(i)}, where the loci of G spiral into the T-point, while S approaches it in a straight line.\footnote{We note that S is located in close proximity of the line of Hopf bifurcations H$^-$, beyond which the simple traveling states are linearly stable. It is possible that the true S crosses the line of H$^-$ at a codimension-2 Shilnikov-Hopf point, and that when a third parameter is varied, the T-point crosses H$^-$ at a codimension-3 T-point-Hopf bifurcation. In these cases, homo- and heteroclinic connections to the traveling states transform into connections to the time-periodic states that emerge subcritically at H$^-$. The one-parameter snaking structure of the shuttling states and the two-parameter spiral of G remain qualitatively similar when crossing these bifurcations \citep{HiKn1993pd, AlFeMeRo2015cnsms}.} We do not resolve this structure but note that in region T we find linearly stable traveling, shuttling, crawling and uniformly covered resting droplets. Lastly, we observe that the shuttling states that emerge from S cannot be observed for low $D$, corresponding to an end of the bifurcation line [region A in Fig.~\subref{fig:bifurcation_diagrams}{(c)}]. Because the Shilnikov condition $\vert \rho/\lambda_3\vert<1$ is still fulfilled in this region, we speculate that here, the line S turns around toward greater $D$. For a discussion of other possible termination mechanisms, see  Ref.~\citep{CKKO2007sjoads}.

\subsection{3D droplets}
\begin{figure}
%\centering
\includegraphics[scale=1]{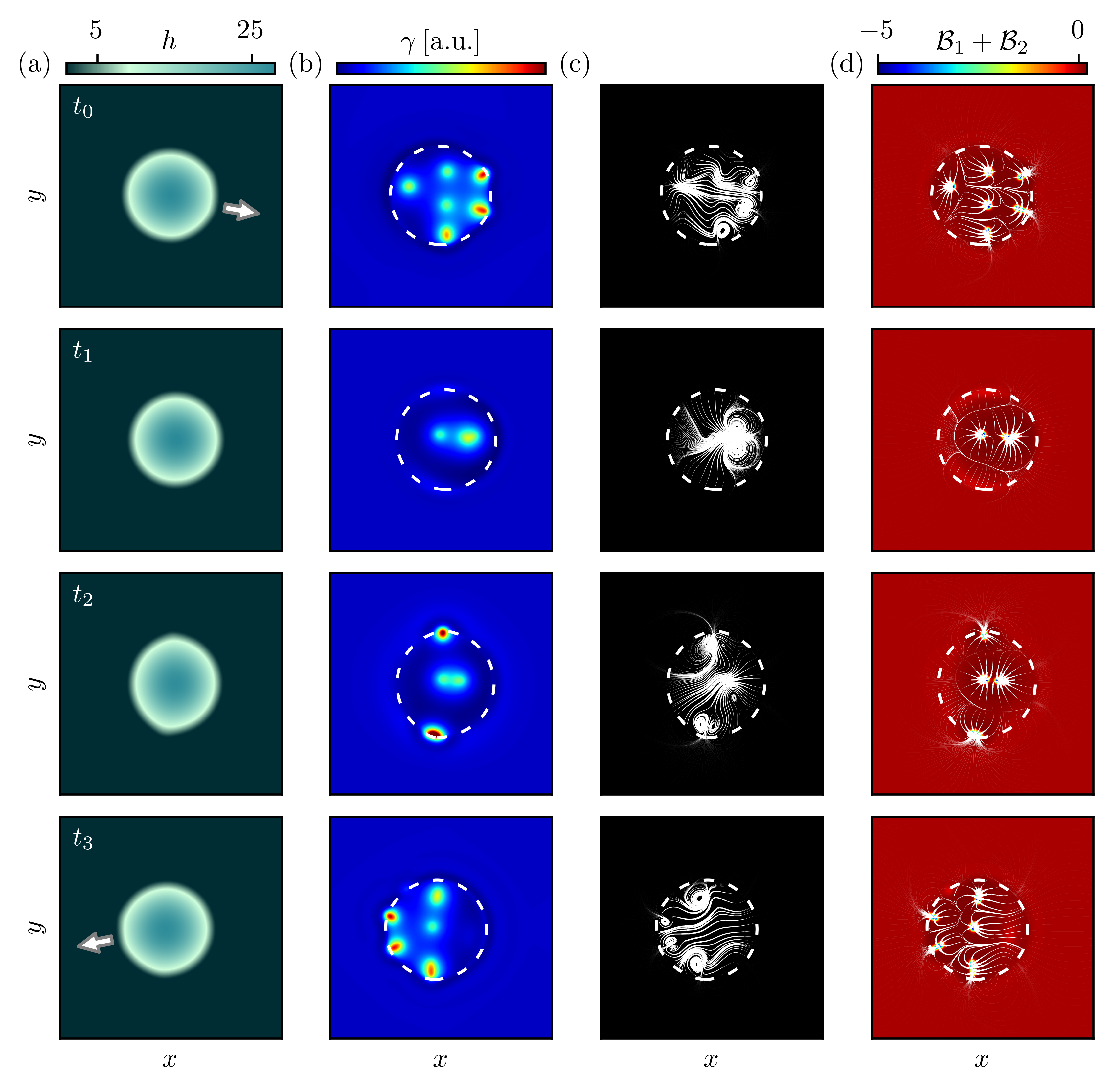}
\caption{Snapshots from a time simulation of a large 3D droplet performing a random walk. The columns show (a) the film thickness, (b) the surface tension, (c) the streamlines of the liquid transport flux $\bm{j}_h$ and (d) both the streamlines of the total surfactant transport flux $\bm{j}_1+\bm{j}_2$ (white lines) and the local total source term of surfactant $\mathcal{B}_1+\mathcal{B}_2$ (color map). The streamline thickness corresponds to the magnitude of the flux. Dashed contour lines in (b), (c) and (d) show the outline of the droplet (given by $h=1.1$). In (b), regions of high surface tension are colored in red. The individual rows are at different times $t_0,\ldots, t_3$ and show a typical change in the direction of motion resulting from the merging of the existing surface tension peaks ($t_0$ and $t_1$) and a subsequent generation of new peaks at different positions ($t_2$ and $t_3$). Note that the streamlines in (c) converge in regions of low $\mathcal{B}_1+\mathcal{B}_2$ that are surfactant sinks. Arrows in (a) indicate the direction of motion. The parameters are $W=10, r=0.4, D_1=D_2=0.2, \beta_1=2, \beta_2=0.01, \mu_1=-1.4, \mu_2=4.175, \delta=1$ with a mean film thickness of $\bar{h}=3$. The periodic computational domain is $\left[0,100\right]\times \left[0,100\right]$. See also Supplemental Video 9.}
\label{fig:2D_chaotic_motion}
\end{figure}

Up to here, we have discussed 2D droplets on 1D substrates. However, our considerations of the feedback loop in Sec.~\ref{sec:simple_traveling} also apply to the case of a 3D drop on a 2D substrate. The liquid transport flux $\bm{j}_h$ and the surfactant transport fluxes $\bm{j}_1$ and $\bm{j}_2$ in  Eqs.~(\ref{eq:model}) are then two-dimensional vector fields and can exhibit,~e.g., vortex structures. This may result in more complex dynamics, particularly in parameter regions where 2D drops show shuttling or crawling motion. The snapshots from a time simulation presented in Fig.~\ref{fig:2D_chaotic_motion} show a typical change of the direction of droplet motion (Supplemental Video 9). Localized surface tension gradients are simultaneously excited in different regions of the droplet, breaking rotational symmetry and leading to complex liquid flow with several vortices near areas of increased surface tension. The Marangoni stresses between such regions then cause fluid flows that result in an effective attraction between neighboring surface tension peaks. Ultimately, this results in an irregular dynamics of generation, attraction and merging of these localized structures. Sufficiently large droplets then no longer move along straight lines but can explore the entire substrate in a random walk. This is mediated by the repeated nucleation of protruding regions of high surface tension at various positions along the contact line. Interestingly, at small liquid volumes shuttling motion is recovered (not shown).

\begin{figure}
%\centering
\includegraphics[scale=1.]{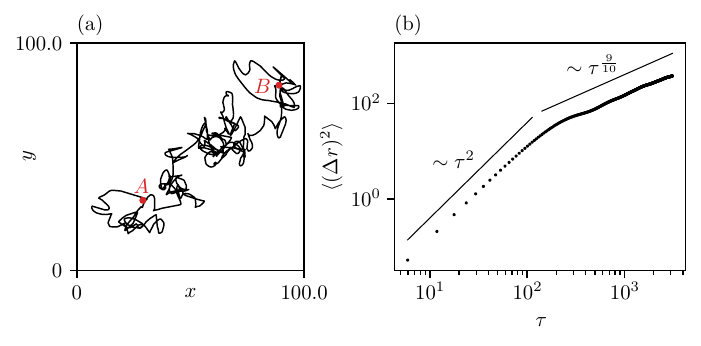}
\caption{(a) Trajectory of the center of mass of a droplet undergoing a random walk from point $A$ to point $B$ (red dots) over an approximate time interval of $t_B\!-\!t_A\!=3\!\times\!10^4$. (b) Corresponding mean square displacement $\langle (\Delta r)^2\rangle$ for the trajectory in (a). There are two scaling regimes with $\langle (\Delta r)^2\rangle \sim \tau^\alpha$. For short times, droplet motion is ballistic with $\alpha\approx 2$ and for longer times it is approximately diffusive with $\alpha\approx 9/10$. The parameters are as in Fig.~\ref{fig:2D_chaotic_motion}.}
\label{fig:trajectory_MSD}
\end{figure}

A typical center-of-mass trajectory of such irregular motion is shown in Fig.~\subref{fig:trajectory_MSD}{(a)}. We characterize this motion in  Fig.~\subref{fig:trajectory_MSD}{(b)} by computing the mean squared displacement $\langle (\Delta r)^2\rangle (\tau) = \frac{1}{t_B-t_A}\int_{t_A}^{t_B}\vert\vert \bm{r}(t+\tau)-\bm{r}(t)\vert\vert^2\:\text{d}t$, where $t_A$ and $t_B$ are the start and end times of the trajectory and $\bm{r}$ denotes the droplet center of mass (Appendix \ref{app:numerical_methods}). We find two scaling regimes characterized by different power laws $\langle (\Delta r)^2\rangle \sim \tau^\alpha$. At short times, we have $\alpha\approx2$,~i.e., the motion is ballistic. At longer times, we have $\alpha\approx 9/10$ and droplet motion is to a good approximation diffusive ($\alpha=1$). Note that the small undulations visible in this regime correspond to decaying oscillations in the velocity auto-correlation function $\langle \bm{v}\cdot \bm{v}\rangle (\tau) = \frac{1}{2}\frac{\text{d}^2\langle (\Delta r)^2\rangle}{\text{d}\tau^2}$ \cite{delC2008}. This implies that droplets are slightly more likely to reverse their direction of travel than to turn in any other direction (for pure shuttling motion one would have a non-decaying oscillation of $\langle \bm{v}\cdot \bm{v}\rangle$). Therefore, the random walk is not entirely uncorrelated. Nevertheless, we expect that the resulting deviations from ideal diffusion are transients, i.e., $\langle (\Delta r)^2\rangle \sim \tau$ should be approached asymptotically as $\tau\rightarrow\infty$.

\section{Conclusion and Outlook}\label{sec:conclusion}
Many biomimetic and biological systems involve chemomechanical coupling with interacting chemical reactions and hydrodynamic transport resulting in complex spatio-temporal behavior. To study such interactions, we have employed a conceptually simple mesoscopic hydrodynamic model that captures the dynamics of sessile droplets of a partially wetting liquid covered by autocatalytically reacting surfactants. The droplets are supplied with chemical fuel by an external chemostat. Notably, our description is based on nonequilibrium thermodynamics, ensuring the existence of a thermodynamically consistent passive (\enquote{dead}) limit. Our study has focused on the self-propulsion of droplets that emerges for sufficiently large nonequilibrium driving. The underlying mechanism is a positive feedback loop between the solutal Marangoni effect and the local chemical reactions. Importantly, in contrast to the commonly treated drop-based microswimmers with simple conversion reactions \cite{MKHB2016arcmp,Mich2023arfm}, the mechanism is not based on differences in physical properties of different surfactants. The considered hydrodynamic scenario is also distinct since we treat droplets on a substrate in a stress-free ambient medium instead of fully immersed drops in a bulk fluid that contributes to the dynamics. Relaxing the latter assumption, one could additionally consider flows in the surrounding medium,~e.g., along the lines of \citep{MPBT2005pf}. Similarly, the present model may be augmented to account for the surfactant dynamics in the bath (which may correspond to either the droplet bulk or the ambient medium), i.e., by treating soluble surfactants \citep{ThAP2016prf} or surfactants diffusing like a vapor phase \citep{HDJT2023jfm}.

Besides the simple self-propelled drops that move uniformly at constant speed, we find that droplets on 1D substrates may also crawl (periodic stick-slip motion) or shuttle (periodic back-and-forth motion). These types of motility can be explained by the positive feedback between Marangoni fluxes and the local reactor. However, in the context of dynamical systems they can also be discussed as periodic orbits in an effective low-dimensional phase space, where the corresponding trajectories closely pass fixed points representing traveling and resting states. Using this representation, we have identified two scenarios involving global bifurcations for the transitions between crawling and shuttling states. Based on two-parameter representations of the loci of a gluing and a Shilnikov global bifurcation in the $(\mu_2, D)$-plane, we have further speculated that both bifurcations terminate at a codimension-2 point (T-point). Various different linearly stable resting and motile droplets exist in the vicinity of this point. It therefore acts as an organizing center in parameter space. Similar motifs arise in the description of biological systems,~e.g., in the study of electrical or mechanical signaling in resting cells \cite{IZHI2000ijobac,DuBAPE2025prl} and stick-slip cycles of motile cells \citep{RMGV2020prr} where small parameter changes allow cells to switch between different modes of operation. For the actin cortex of motile cells, it has been suggested on the basis of mass-conserving reaction-diffusion models that the bistability of various dynamical states of the cortex can be traced to codimension-2 points \citep{YoFB2022prl, HuMEY2024}. Here, we find the organization of highly complex states of motility around a codimension-2 point for a full, spatially extended dynamical model that couples various transport and reaction processes in the presence of interfaces, which to our knowledge has not yet been reported. 

In our study of shuttling and crawling states, we have only employed direct numerical simulations since numerical continuation of such spatially localized time-periodic slow-fast states is numerically exceptionally challenging and computationally expensive. In consequence, we have only been able to obtain partial bifurcation diagrams without the unstable branch sections of crawling and shuttling droplets. In the future, our results could be extended by studying the entire bifurcation structure,~e.g., by employing spatio-temporally adaptive continuation methods. One could thereby gain information about local bifurcations of such states. For example, we expect that branches of crawling states terminate either at Hopf bifurcations of traveling drops or at drift-pitchfork bifurcations of resting oscillating drops. Further, it is known that direction-reversing (shuttling) states can emerge in reflection symmetry-breaking Hopf bifurcations \citep{TaOh2016el,OKGT2020c,LaKn1991pla}. To verify our hypothesis that the two-parameter bifurcation diagram is organized about a T-point, it is necessary to employ numerical continuation of the associated heteroclinic and homoclinic connections involving the resting and traveling states - a task beyond present numerical tools for partial differential equations known to us. Lastly, we have turned to 3D droplets that, when large enough, explore the entire available substrate in an approximately diffusive random walk due to irregular excitation, attraction and merging of localized spots of surface tension gradients near the contact line region. The transition between such random walks and other motility modes shall be investigated in the future. Here, we expect that particularly the droplet volume represents an important control parameter, since larger droplets can accommodate a larger number of localized surface tension peaks, resulting in more complex behavior. This holds for both 2D and 3D drops, i.e., we find that large 2D droplets ($\sim 50\%$ larger volume than the one used in Sections \ref{sec:onset_selfpropulsion}-\ref{sec:higher_codim}) also exhibit highly irregular dynamics. Furthermore, the linear stability of the traveling and resting states may change with the drop volume (e.g., the symmetric one-peak resting state may become linearly stable for small volumes), thus altering the bifurcation picture. In this regard, we expect that the results presented in Sec.~\ref{sec:bifurcation_analysis} are valid for drops that can accommodate one or two surface tension peaks.

The discussed positive feedback loop is likely relevant to many biomimetic droplet systems that feature higher-order chemical reactions of surface active species \citep{{MaYo1996jpc,SMAN2016jpcl,TaSN2015pre,SSK2017sm, TaNN2021sm}}. Such reactions may occur,~e.g., in auto- or cross-catalytic mechanisms and the formation of micellar structures. Moreover, while the presented simple model does not reflect the complex biochemistry of real biological cells, it may capture a motility mechanism relevant under prebiotic conditions,~i.e., for protocells, where autocatalytic reactions may represent a simple form of molecular self-replication. Additionally, the here studied forms of motility (crawling, shuttling with and without net drift, irregular motion) appear in models and experiments for amoeba \citep{DrAK2014njp} and Physarum droplets \citep{RoTUH2015ebj,LeZGA2015jrsi, KuJBE2019po}. Although related theoretical descriptions may include mechanical effects that go beyond the present hydrodynamic model (e.g., viscoelasticity), it is worth pointing out that they typically feature a (linear) coupling of a nonlinear chemical kinetics and active mechanical stresses, in analogy to the here considered coupling of surface tension effects and autocatalytic chemical reactions via surfactants. Furthermore, because common recent descriptions of biomolecular condensates within cells are based on a similar thermodynamic structure \cite{GDMF2024prr,DGMF2023prl,WZJL2019rpp,Zwic2022cocis}, we believe that also there similar states and higher-codimension organizing centers may be of importance if one includes cross- or autocatalytic effects and considers nonlinear mobility functions. In our case, the latter result from advection but they may also arise,~e.g., for diffusion in crowded environments \citep{RHZJ2011pnas}. Finally, we remark that one may lift our present restriction to the case where the underlying reaction system does not form any patterns. Then one needs to amend the mobilities and reaction rates,~e.g., by adding film thickness-dependent cutoffs of the rates to maintain a passive adsorption layer representing a macroscopically dry and therefore passive substrate.

\appendix
\section{Dynamical equations in hydrodynamic form}\label{app:hydrodynamic_form}
Equations~(\ref{eq:model}) can be expressed in hydrodynamic form. To this end, we first explicitly compute the variations
\begin{equation}
\begin{aligned}
p &= \frac{\delta F}{\delta h} = \partial_h f- \nabla\cdot\left[\frac{1}{\xi}\left(g-\Gamma_1\partial_{\Gamma_1}g-\Gamma_2\partial_{\Gamma_2}g\right)\nabla h\right],\\
\mu_\alpha &= \frac{\delta F}{\delta \hat{\Gamma}_\alpha} =\partial_{\Gamma_\alpha}g \hspace*{0.5cm}\text{with}\hspace*{0.5cm}\alpha=1,2.
\end{aligned}
\label{app:eq:_p_mu}
\end{equation}
In particular, with the energetic contribution (\ref{eq:surface_energy_definition}) for $g(\Gamma_1,\Gamma_2)$ we obtain
\begin{equation}
\begin{aligned}
p &= \partial_h f - \nabla\cdot\left[\frac{1}{\xi}\gamma(\Gamma_1,\Gamma_2)\nabla h\right]\\
\mu_\alpha &= k_bT\ln(\Gamma_\alpha a_\alpha^2)\hspace*{0.5cm}\text{with}\hspace*{0.5cm}\alpha=1,2.
\end{aligned}
\label{app:eq:p_mu_substituted}
\end{equation}
Here, $\gamma(\Gamma_1, \Gamma_2) = \gamma_0-k_bT(\Gamma_1+\Gamma_2)$ is the resulting local surface tension. By substituting the expression for the chemical potentials in Eqs.~(\ref{app:eq:p_mu_substituted}) into Eqs.~(\ref{eq:model}) and transitioning to the long-wave limit $\vert \nabla h\vert\ll 1$ (yielding $\hat{\Gamma}_\alpha \approx \Gamma_\alpha$), one obtains the hydrodynamic form
\begin{equation}
\begin{aligned}
\partial_t h &= \nabla\cdot\left[\frac{h^3}{3\eta}\nabla p -\frac{h^2}{2\eta}\nabla \gamma \right],\\
\partial_t \Gamma_1 &= \nabla\cdot\left[\frac{h^2\Gamma_1}{2\eta}\nabla p-\frac{h\Gamma_1}{\eta}\nabla \gamma \right]+D_1k_bT\Delta\Gamma_1+\mathcal{R}+\mathcal{B}_1,\\
\partial_t \Gamma_2 &= \nabla\cdot\left[\frac{h^2\Gamma_2}{2\eta}\nabla p-\frac{h\Gamma_2}{\eta}\nabla \gamma \right]+D_2k_bT\Delta\Gamma_2-\mathcal{R}+\mathcal{B}_2
\end{aligned}
\label{app:eq:hydrodyn_model}
\end{equation}
where $\mathcal{R}, \mathcal{B}_1$ and $\mathcal{B}_2$ are given by Eqs.~(\ref{eq:autocatalysis_detailed_balanced}) and (\ref{eq:adsorption_desorption}). Further, it is commonly assumed that the changes in surface tension due to the surfactants are small compared to $\gamma_0$ \citep{CrMa2009rmp}. The pressure then reduces to $p=\partial_h f -\gamma_0\Delta h$. This approximation only affects the capillary pressure and does not alter the Marangoni fluxes appearing in Eqs.~(\ref{app:eq:hydrodyn_model}). Note that if one also transitions to the long-wave limit as done here, one again obtains a gradient dynamics (for $\mu_1=\mu_2$) on the simplified grand potential $\Omega^*=\int_\mathcal{S}\left[f(h)+\frac{\gamma_0}{2}\vert\nabla h\vert^2+g(\Gamma_1,\Gamma_2)-\sum_i\mu_i\Gamma_i\right]\text{d}^2x$, where the mobility matrix is given by Eq.~(\ref{eq:autocatalyic_surfactants_Q}). The reaction currents (\ref{eq:autocatalysis_detailed_balanced}) and (\ref{eq:adsorption_desorption}) can then be re-expressed as
\begin{equation}
\begin{aligned}
\mathcal{R}&=r^*\left[\exp\left(\frac{2}{k_b T}\frac{\delta \Omega^*}{\delta \Gamma_1}+\frac{1}{k_b T}\frac{\delta \Omega^*}{\delta \Gamma_2}\right)-\exp\left(\frac{3}{k_b T}\frac{\delta \Omega^*}{\delta {\Gamma}_1}\right)\right],\\
\mathcal{B}_1&=\beta^*_{1}\left[1-\exp\left(\frac{1}{k_b T}\frac{\delta \Omega^*}{\delta \Gamma_{1}}\right)\right],\\
\mathcal{B}_2&=\beta^*_{2}\left[1-\exp\left(\frac{1}{k_b T}\frac{\delta \Omega^*}{\delta \Gamma_{2}}\right)\right],
\end{aligned}
\label{app:R_B12_grand_pot}
\end{equation}
with $r^*=re^{\frac{3\mu_1}{k_bT}}, \beta_1^*=\beta_1e^{\frac{\mu_1}{k_bT}}$ and $\beta_2^*=\beta_2e^{\frac{\mu_1}{k_bT}}$ as the new rate constants. Then, all currents in (\ref{app:R_B12_grand_pot}) lead to a monotonic decrease of $\Omega^*$,~i.e., they are purely dissipative.

\section{Nondimensionalization}\label{app:nondimensionalization}
We rescale Eqs.~(\ref{app:eq:hydrodyn_model}) by introducing the scalings
\begin{equation}
t = \tau\tilde{t}, \hspace*{0.5cm} (x,y)=L(\tilde{x}, \tilde{y}), \hspace*{0.5cm} h=l\tilde{h} , \hspace*{0.5cm} \Gamma_\alpha=\tilde{\Gamma}_\alpha/(a_1a_2),  \hspace*{0.5cm} (f, g) = \kappa(\tilde{f}, \tilde{g}), \label{app:eq:nondimensionalization}
\end{equation}
where $\alpha=1,2$ and dimensionless quantities are denoted by a tilde. The scales are chosen as
\begin{equation}
\tau = \frac{L^2\eta}{\kappa l}, \hspace*{0.5cm} L=\sqrt{\frac{\gamma_0}{\kappa}}l, \hspace*{0.5cm} l=h_a,  \hspace*{0.5cm} \kappa=\frac{k_b T}{a_1 a_2}. \label{app:eq:scaling}
\end{equation}
Note that the long-wave limit $L\gg l$ implies $\gamma_0\gg\frac{k_bT}{a_1a_2}$ in the present scaling, which is therefore consistent with the approximation $p=\partial_h f -\gamma_0\Delta h$ as outlined in Appendix~\ref{app:hydrodynamic_form}. This yields the nondimensional parameters
\begin{equation}
\begin{aligned}
\delta &= \frac{a_1}{a_2},\\
W &= \frac{A}{l^2\kappa}, \\
\tilde{D}_\alpha &= \frac{\tau}{L^2}k_bT D_\alpha, \\
\tilde{r} &= \tau a_1a_2 r, \\
\tilde{\beta}_\alpha &= \tau a_1a_2 \beta_\alpha,\\
\tilde{\mu}_\alpha &= \frac{1}{k_bT}\mu_\alpha,
\end{aligned} \label{app:eq:nondimensional_parameters}
\end{equation}
with $\alpha=1,2$. Omitting tildes, we obtain the dimensionless equations
\begin{equation}
\begin{aligned}
\partial_t h &= \nabla\cdot\left[\frac{h^3}{3}\nabla p +\frac{h^2}{2}\nabla (\Gamma_1+\Gamma_2)\right],\\
\partial_t \Gamma_1 &= \nabla\cdot\left[\frac{h^2\Gamma_1}{2}\nabla p+h\Gamma_1\nabla (\Gamma_1+\Gamma_2) \right]+D_1\Delta\Gamma_1+\mathcal{R}+\mathcal{B}_1,\\
\partial_t \Gamma_2 &= \nabla\cdot\left[\frac{h^2\Gamma_2}{2\eta}\nabla p+h\Gamma_2\nabla (\Gamma_1+\Gamma_2) \right]+D_2\Delta\Gamma_2-\mathcal{R}+\mathcal{B}_2,
\end{aligned}
\label{app:eq:model_nondim}
\end{equation}
where the (simplified) pressure is given by
\begin{equation}
p = W\left(\frac{1}{h^3}-\frac{1}{h^6}\right)-\Delta h
\end{equation}
and the reaction terms are
\begin{equation}
\begin{aligned}
\mathcal{R} &= r\left[\delta\Gamma_2\Gamma_1^2-(\delta\Gamma_1)^3\right], \\
\mathcal{B}_1&=\beta_1\left[e^{\mu_1}-\delta\Gamma_1\right], \\
\mathcal{B}_2&=\beta_2\left[e^{\mu_2}-\delta^{-1}\Gamma_2\right].
\end{aligned}
\end{equation}
Table \ref{app:tab:dimensional_quantities_example} lists a possible set of dimensional parameters, resulting in the nondimensional parameter choice of Fig.~\subref{fig:selfpropelled_droplets_overview}{(a)}. We use the given values of $l, a_1, a_2, k_bT, \eta$ and $\gamma_0$ to determine the scales (\ref{app:eq:scaling}) and then the other dimensional parameters using (\ref{app:eq:nondimensional_parameters}). Here, the choice of $l$ yields micrometer-sized droplets (height $\sim\SI{10}{\micro\meter}$, diameter $\sim\SI{250}{\micro\meter})$. For the surfactant length scales, we follow Ref.~\citep{TrJT2018sm}. The system is assumed to be at room temperature ($T=\SI{25}{\celsius}$) and we assume a surface tension $\gamma_0$ of the bare droplet (without surfactant) that is comparable to that of the water-air interface. For the dynamic viscosity $\eta$, we assume that the bulk liquid is significantly more viscous than water but within the typical range encountered for applications concerning thin films containing surfactant \citep{CrMa2009rmp}. Notably, this results in a time scale of $\tau\approx \SI{0.03}{\second}$, yielding typical shuttling periods of $\sim \SI{20}{\second}$. Similar values have been reported experimentally for the shuttling motion of droplets in the presence of multiple (possibly nonlinearly) chemically reacting surfactants \citep{SSK2017sm,TaNN2021sm}. Further, the dimensional surfactant surface diffusivities are also consistent with literature values \citep{SaBe1969iecf, Ni2003bj}. Nevertheless, we stress that the aim of the the presented study is not a quantitative comparison with experiments but the qualitative investigation of the interplay between (nonlinear) chemical reactions on interfaces and droplet hydrodynamics.
\begin{table} 
\caption{Example set of approximate numerical values for the scales employed in the nondimensionalization (\ref{app:eq:nondimensionalization}) and the dimensional parameters of the model (\ref{eq:model}).}
\begin{tabular}{cccc}
\toprule
Symbol & Quantity & Numerical Value \\  \midrule 
$l$ & vertical length scale & \SI{0.5}{\micro\meter}\\ 
$a_1 ,a_2$ & surfactant length scales & \SI{3}{\nano\meter} \\
$k_bT$ & thermal energy (at \SI{25}{\celsius})& \SI{4e-21}{\joule} \\
$\eta$ & dynamic viscosity & \SI{0.2}{\pascal\second} \\
$\gamma_0$ & surface tension (bare droplet) & \SI{70e-3}{\joule\per\meter\squared}\\
$\kappa$ & thermal energy density of surfactant  & \SI{4e-4}{\joule\per\meter\squared}\\
$L$ & horizontal length scale & \SI{5}{\micro\meter}\\ 
$\tau$ & time scale & \SI{0.03}{\second}\\ 
$A$ & Hamaker constant & \SI{1e-15}{\joule}\\ 
$\theta_{eq}$ & equilibrium contact angle (bare droplet) & \SI{10}{\degree}\\
$\frac{L^2}{\tau}\tilde{D}_{1},\frac{L^2}{\tau}\tilde{D}_2$ & surfactant diffusivity (on interface) & \SI{4e-10}{\meter\squared\per\second}\\
$\frac{1}{\tau}\tilde{r},\frac{1}{\tau}\tilde{\beta}_1,\frac{1}{\tau}\tilde{\beta}_2$ & reaction rate constants & \SI{10}{\per\second}, \SI{0.3}{\per\second},\SI{70}{\per\second}\\
$\mu_1, \mu_2$ & chemical potentials (chemostats) & -1\;$k_bT$, 4\;$k_bT$ \\
\bottomrule
\end{tabular} 
\label{app:tab:dimensional_quantities_example}
\end{table}

\section{Hopf bifurcations of the local reactor}\label{app:local_reactor}
To obtain condition (\ref{eq:reactor_hopf_bifs}), we consider the Jacobian of Eqs.~(\ref{eq:local_reactor}) at the fixed point
\begin{equation}
\tens{J} =	\left(\begin{array}{cc}
\partial_{\Gamma_1}\mathcal{R}+\partial_{\Gamma_1}\mathcal{B}_1 & \partial_{\Gamma_2}\mathcal{R}\\ 
-\partial_{\Gamma_1}\mathcal{R}  & -\partial_{\Gamma_2}\mathcal{R}+\partial_{\Gamma_2}\mathcal{B}_2
\end{array} \right)_{\Gamma_{1,ss},\Gamma_{2,ss}}.
\end{equation}
When the fixed point is located near the extrema of the $\Gamma_1$-nullcline both nullclines are approximately perpendicular and locally align with the coordinate axes (the nullcline of $\Gamma_1$ aligns with the $\Gamma_1$-axis and the nullcline of $\Gamma_2$ with the $\Gamma_2$-axis). The flow of Eqs.~(\ref{eq:local_reactor}) is then circular near the fixed point and the eigenvalues are complex. Hopf bifurcations are therefore given by the condition $\tr\tens{J}\!=\!0$. Using $\partial_{\Gamma_2}\mathcal{B}_2=\delta^{-1}\beta_2=\mathcal{O}(\varepsilon)$, this requirement simplifies at $\mathcal{O}(1)$ to
\begin{equation}
0=r\left[2\delta \Gamma_{2,ss}\Gamma_{1,ss}-(3\delta^3+\delta)\Gamma_{1,ss}^2\right]-\delta\beta_1. \label{app:eq:trace_simplified}
\end{equation}
Using Eqs.~(\ref{eq:reactor_fp}), we obtain from relation (\ref{app:eq:trace_simplified})
\begin{equation}
0=(\delta+\delta^3)r\Gamma_{1,ss}^2+\frac{2}{\Gamma_{1,ss}}\beta_1 e^{\mu_1}-\delta\beta_1. \label{app:eq:hopf_condition}
\end{equation}
Furthermore, Eq.~(\ref{eq:reactor_hopf_bifs}) can be re-expressed using the expression for $\Gamma_{1,ss}$ in (\ref{eq:reactor_fp}) and the chain rule,
\begin{equation}
\frac{\partial\left(\Gamma_{1,ss}+\Gamma_{2,ss}\right)}{\partial \Gamma_{1,ss}}=0,\label{app:eq:surface_tension_extrema_chain_rule}
\end{equation}
where $\Gamma_{2,ss}$ is understood as a function of $\Gamma_{1,ss}$ and of all other parameters (except $\beta_2$ and $\mu_2$) appearing in Eqs.~(\ref{eq:reactor_fp}). By using Eqs.~(\ref{eq:reactor_fp}) one also obtains the relation $\Gamma_{2,ss}=\delta^2\Gamma_{1,ss}+\frac{1}{r\delta\Gamma_{1,ss}^2}\left(\delta\beta_1\Gamma_{1,ss}-\beta_1e^{\mu_1}\right)$. Substituting this into (\ref{app:eq:surface_tension_extrema_chain_rule}) again yields (\ref{app:eq:hopf_condition}).

\section{Approximations for flux-reaction balance}\label{app:approximations_flux_reaction_balance}
\begin{figure}
\centering
\includegraphics[scale=1.1]{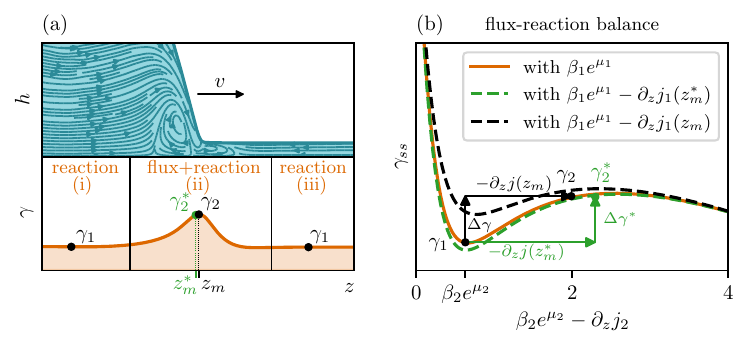}
\caption{(a) Magnification of the advancing contact line region of a self-propelled droplet in the comoving frame $z=x-vt$. The top panel shows the film-height profile and the velocity field of the liquid (in the laboratory frame), the bottom panel shows the surface tension profile. The two points $z_m$ and $z_m^*$ are the surface tension maximum ($\gamma=\gamma_2$) and a point slightly offset from it ($\gamma=\gamma_2^*$). (b) Steady state curve $\gamma_{ss}$ of the local reactor as a function of $\beta_2e^{\mu_2}-\partial_zj_2$. We consider three values of the secondary parameter (see text), corresponding to $\beta_1e^{\mu_1}$ (solid orange line), $\beta_1e^{\mu_1}-\partial_zj_1(z_m)$ (dashed black line) and $\beta_1e^{\mu_1}-\partial_zj_1(z_m^*)$ (dashed green line). When considering $\gamma_2$, there are significant deviations between the cases $\beta_1e^{\mu_1}$ and $\beta_1e^{\mu_1}-\partial_zj_1(z_m)$. For $\gamma_2^*$, the cases $\beta_1e^{\mu_1}$ and $\beta_1e^{\mu_1}-\partial_zj_1(z_m^*)$ nearly overlap and $\gamma_2^*$ approximately lies on both curves.}
\label{app:fig:flux_reaction_balance}
\end{figure}

During our discussion of the self-propulsion mechanism in Sec.~\ref{sec:simple_traveling}, we have made two simplifying assumptions. Generally, for a steadily traveling droplet in the comoving frame, the surfactant profiles are given by
\begin{equation}
\begin{aligned}
0&=-\partial_z j_1+\mathcal{R}+\mathcal{B}_1-v\partial_z\Gamma_1,\\
0&=-\partial_z j_2-\mathcal{R}+\mathcal{B}_2-v\partial_z\Gamma_2.
\end{aligned} 
\label{app:eq:comoving_frame}
\end{equation}
First, we have assumed that the extrema of $\Gamma_1, \Gamma_2$ and $\gamma$ coincide in space. This allows us to neglect the terms $-v\partial_z\Gamma_1,-v\partial_z\Gamma_2$ at the surface tension peak ($z=z_m$) in region (ii) [see Fig.~\subref{app:fig:flux_reaction_balance}{(a)}], yielding the balance equations
\begin{equation}
\begin{aligned}
0&=-\partial_z j_1+\mathcal{R}+\mathcal{B}_1, \\
0&=-\partial_z j_2-\mathcal{R}+\mathcal{B}_2,
\end{aligned}
\label{app:eq:comoving_frame_no_adv}
\end{equation}
at $z=z_m$ in the comoving frame. Second, we have argued that the transport contribution $-\partial_z j_1$ is negligible at the peak since diffusive and advective contributions effectively cancel. One then obtains
\begin{equation}
\begin{aligned}
0&=\mathcal{R}+\mathcal{B}_1,\\
0&=-\partial_z j_2-\mathcal{R}+\mathcal{B}_2,
\end{aligned}
\label{app:eq:comoving_frame_simplified} 
\end{equation}
As described in the main text, Eqs.~(\ref{app:eq:comoving_frame_simplified}) correspond to the steady state equations of the local reactor with the shifted parameter $\beta_2e^{\mu_2}-\partial_zj_2(z_m)$,~i.e., the surface tension at $z_m$ can be obtained by evaluating $\gamma_{ss}(\beta_2e^{\mu_2}-\partial_zj_2(z_m))$. Analogously, if $-\partial_z j_1$ is not neglected, one may obtain the peak surface tension by shifting both driving parameters,~i.e., by evaluating $\gamma_{ss}(\beta_2e^{\mu_2}-\partial_z j_2(z_m),\beta_1e^{\mu_1}-\partial_z j_1(z_m))$. For the local reactor, this corresponds to considering the function $\gamma_{ss}(\beta_2e^{\mu_2}, \beta_1e^{\mu_1})$ instead, where $\beta_1e^{\mu_1}$ is a secondary parameter [cf.~Fig.~\subref{fig:ode_summary}{(f)}]. Although we justify dropping the contribution $-\partial_z j_1(z_m)$ in the following, we point out that doing so is not strictly necessary. Rather, it simplifies the discussion in Sec.~\ref{sec:simple_traveling}. From Fig.~\subref{fig:ode_summary}{(f)} it can be seen that changes in $\beta_1 e^{\mu_1}$ [here $\beta_1e^{\mu_1}-\partial_z j_1$] only influence the shape of $\gamma_{ss}$ as a function of $\beta_2 e^{\mu_2}$ [here $\beta_2e^{\mu_2}-\partial_z j_2$] near its local minimum, while the curve remains mostly unaffected for larger values of $\beta_2 e^{\mu_2}$ near its local maximum. It is this section of $\gamma_{ss}$ near its local maximum that is relevant for points in space near the localized surface tension peak at $z=z_m$. Therefore, using Eq.~(\ref{app:eq:comoving_frame_no_adv}) instead of (\ref{app:eq:comoving_frame_simplified}) only marginally affects $\gamma_2$.

Nevertheless, we generally find that Eqs.~(\ref{app:eq:comoving_frame_simplified}) are appropriate to describe $\gamma_2$,~i.e., it indeed holds that $\gamma_2 = \gamma_{ss}(\beta_2e^{\mu_2}-\partial_zj_2(z_m))$ [Figs.~\subref{fig:flux_reaction_balance}{(b)} and \subref{app:fig:flux_reaction_balance}{(b)}]. However, a caveat is in order. By comparing the curves $\gamma_{ss}$ as a function of $\beta_2e^{\mu_2}-\partial_z j_2$ either with or without shifting $\beta_1e^{\mu_1}$ (the steady state surface tension corresponding to either Eqs.~(\ref{app:eq:comoving_frame_no_adv}) or (\ref{app:eq:comoving_frame_simplified}), respectively) we find that these curves differ significantly,~i.e., the contribution $-\partial_zj_1(z_m)$ is not negligible [Fig.~\subref{app:fig:flux_reaction_balance}{(b)}]. The reason is that the terms $-v\partial_z\Gamma_1,-v\partial_z\Gamma_2$ and $-\partial_zj_1$ do not vanish at the same point in space (if they did, our approximations would be exact) and the corresponding contributions in Eqs.~(\ref{app:eq:comoving_frame}) are large enough to significantly change $\gamma_{ss}$ when absorbed into the parameters of the local reactor. The circumstance that $\gamma_2$ is nevertheless well captured by Eqs.~(\ref{app:eq:comoving_frame_simplified}) is in fact a consequence of a favorable cancellation of the errors induced by transitioning from Eqs.~(\ref{app:eq:comoving_frame}) to (\ref{app:eq:comoving_frame_no_adv}) and finally to (\ref{app:eq:comoving_frame_simplified}). However, we find that in practice [for diffusion coefficients ${D=D_1=D_2=\mathcal{O}(10^{-1})}$] one can always find a point $z_m^*$ in the immediate vicinity of $z_m$ with comparable surface tension $\gamma_2^*$ where all three terms $-v\partial_z\Gamma_1(z_m^*), -v\partial_z\Gamma_2(z_m^*)$ and $-\partial_zj_1(z_m^*)$ are indeed negligible (Fig.~\ref{app:fig:flux_reaction_balance}). Therefore, the Eqs.~(\ref{app:eq:comoving_frame_simplified}) generally hold for some point near the surface tension peak with a surface tension that is comparable to the maximum value. This point is usually not $\textit{exactly}$ the surface tension maximum. Because the differences $\Delta\gamma=\gamma_2-\gamma_1$ and $\Delta\gamma^*=\gamma_2^*-\gamma_1$ are nearly identical, we may then draw the same conclusions regarding the interplay between the Marangoni effect and the local reactor as in Sec.~\ref{sec:simple_traveling}. 
 
\section{Numerical Methods}\label{app:numerical_methods}
Direct numerical simulations are based on the finite-element method and are implemented using the open source library oomph-lib \cite{HeHa2006}. Most simulations are performed on a periodic domain that is discretized using a static spatial mesh (1D: 641 nodes, 2D: $251 \times 251$ nodes). An exception are the data shown in Figs.~\subref{fig:flux_reaction_balance}{(a)}, \subref{fig:contact_angle_trajectories}{(a)},\subref{fig:contact_angle_trajectories}{(c)}, \subref{fig:period_scalings}{(a)} and \subref{app:fig:flux_reaction_balance}{(a)}, where we use an adaptive spatial mesh. For temporal discretization, we use a backward differentiation scheme of order 2 (BDF(2)) with adaptive time stepping.

For numerical continuation \citep{KrauskopfOsingaGalan-Vioque2007, DWCD2014ccp,EGUW2019springer, Uecker2021}, Eqs.~(\ref{app:eq:model_nondim}) are spatially discretized using the finite element method on a one-dimensional periodic domain (400 nodes). We employ the continuation package pde2path \citep{Uecker2021} which uses pseudo-arc length continuation with a predictor-corrector method. Because Eqs.~(\ref{app:eq:model_nondim}) exhibit continuous symmetries (liquid volume conservation, translational symmetry on a periodic domain), they are augmented by algebraic constraints for the liquid volume and the spatial phase. Then, two additional parameters must be freed which are also determined during continuation. In particular, the parameter corresponding to the spatial phase constraint is the velocity $v$ of the comoving frame,~i.e., the droplet speed [Fig.~\subref{fig:bifurcation_diagrams}{(d)}]. For details, see Refs.~\citep{EGUW2019springer, Uecker2021}.

The contact angles $\theta_{r,l}$ of shuttling and crawling droplets shown in Figs.~\subref{fig:contact_angle_trajectories}{(a)}-\subref{fig:contact_angle_trajectories}{(c)} are determined from the slopes $m_{r,l}$ of the film-height profiles at the inflection points as $\theta_{r,l}=\atan \vert m_{r,l}\vert$. The trajectories are then smoothed using a Savitzky-Golay filter (window length 53, polynomial order 3, $\sim 1000$ data points per trajectory) using \texttt{scipy.signal.savgol\_filter()} from the Python library SciPy. 

To compute the droplet center of mass on a periodic domain [cf.~Fig.~\subref{fig:trajectory_MSD}{(a)}], we use the algorithm described in Ref.~\citep{BaBre2008}. The mean square displacement $\langle (\Delta r)^2\rangle$ [Fig.~\subref{fig:trajectory_MSD}{(b)}] is obtained in a straight-forward manner,~i.e., by averaging the squared displacement over the discretized center-of-mass trajectory for various temporal shifts. To this end, we first interpolate the center-of-mass trajectory onto a uniform temporal mesh.

\section*{Acknowledgements}
Part of the calculations for this publication were performed on the HPC cluster PALMA II of the University of Münster, subsidized by the Deutsche Forschungsgemeinschaft (DFG) (INST 211/667-1). We acknowledge financial support by the DPG via the grant no. TH 781/12-2 within SPP 2171. FV further acknowledges valuable discussions with Yutaka Sumino and wishes to thank the entire Sumino Lab for their hospitality during his stay in Japan. UT would like to thank the Kavli Institute for Theoretical Physics (KITP), Santa Barbara (grant NSF PHY-2309135) for support and hospitality during the programme \textit{Active Solids} where part of the work was undertaken.

\section*{Conflict of interest}
The authors have no conflicts of interest to disclose.
\section*{Data availability}
The data and python codes used in this study as well as the supplemental videos referenced in the main text are publicly available on Zenodo \citep{VoTh2025zn}.
\section*{Author Contributions}
\textbf{Florian Voss:} Conceptualization (equal); Methodology (equal); Formal analysis (lead); Investigation (lead); Data curation (lead); Software (lead); Visualization (lead); Writing – original draft (equal); Writing – review \& editing (equal). \textbf{Uwe Thiele:} Conceptualization (equal); Methodology (equal); Funding acquisition (lead); Project administration (lead); Supervision (lead); Writing – original draft (equal); Writing – review \& editing (equal).


\begin{thebibliography}{124}
\providecommand{\natexlab}[1]{#1}
\providecommand{\url}[1]{\texttt{#1}}
\expandafter\ifx\csname urlstyle\endcsname\relax
  \providecommand{\doi}[1]{doi: #1}\else
  \providecommand{\doi}{doi: \begingroup \urlstyle{rm}\Url}\fi

\bibitem[Barge et~al.(2015)Barge, Cardoso, Cartwright, Cooper, Cronin, De~Wit,
  Doloboff, Escribano, Goldstein, Haudin, Jones, Mackay, Maselko, Pagano,
  Pantaleone, Russell, Sainz-Diaz, Steinbock, Stone, Tanimoto, and
  Thomas]{BCCC2015cr}
L.~M. Barge, S.~S.~S. Cardoso, J.~H.~E. Cartwright, G.~J.~T. Cooper, L.~Cronin,
  A.~De~Wit, I.~J. Doloboff, B.~Escribano, R.~E. Goldstein, F.~Haudin, D.~E.~H.
  Jones, A.~L. Mackay, J.~Maselko, J.~J. Pagano, J.~Pantaleone, M.~J. Russell,
  C.~I. Sainz-Diaz, O.~Steinbock, D.~A. Stone, Y.~Tanimoto, and N.~L. Thomas.
\newblock From chemical gardens to chemobrionics.
\newblock \emph{Chem. Rev.}, 115:\penalty0 8652--8703, 2015.
\newblock \doi{10.1021/acs.chemrev.5b00014}.

\bibitem[Golestanian et~al.(2005)Golestanian, Liverpool, and
  Ajdari]{GoLA2005prl}
R.~Golestanian, T.~B. Liverpool, and A.~Ajdari.
\newblock Propulsion of a molecular machine by asymmetric distribution of
  reaction products.
\newblock \emph{Phys. Rev. Lett.}, 94:\penalty0 220801, 2005.
\newblock \doi{10.1103/physrevlett.94.220801}.

\bibitem[Sumino et~al.(2023)Sumino, Yamashita, Miyaji, Ishikawa, Otani,
  Yamamoto, Okita, Okamoto, Krafft, Yoshikawa, and Shioi]{SYM2023sr}
Y.~Sumino, R.~Yamashita, K.~Miyaji, H.~Ishikawa, M.~Otani, D.~Yamamoto,
  E.~Okita, Y.~Okamoto, M.~P. Krafft, K.~Yoshikawa, and A.~Shioi.
\newblock Droplet duos on water display pairing, autonomous motion, and
  periodic eruption.
\newblock \emph{Sci Rep}, 13:\penalty0 12377, 2023.
\newblock \doi{10.1038/s41598-023-39094-6}.

\bibitem[Gross et~al.(2017)Gross, Kumar, and Grill]{GrKG2017arb}
P.~Gross, K.~V. Kumar, and S.~W. Grill.
\newblock How active mechanics and regulatory biochemistry combine to form
  patterns in development.
\newblock \emph{Annu. Rev. Biophys.}, 46:\penalty0 337--356, 2017.
\newblock \doi{10.1146/annurev-biophys-070816-033602}.

\bibitem[Maheshwari et~al.(2019)Maheshwari, Sunol, Gonzalez, Endy, and
  Zia]{MSGE2019prf}
A.~J. Maheshwari, A.~M. Sunol, E.~Gonzalez, D.~Endy, and R.~N. Zia.
\newblock Colloidal hydrodynamics of biological cells: A frontier spanning two
  fields.
\newblock \emph{Phys. Rev. Fluids}, 4:\penalty0 110506, 2019.
\newblock \doi{10.1103/physrevfluids.4.110506}.

\bibitem[Rafelski and Theriot(2004)]{RaTh2004arb}
S.~M. Rafelski and J.~A. Theriot.
\newblock Crawling toward a unified model of cell motility: Spatial and
  temporal regulation of actin dynamics.
\newblock \emph{Annu. Rev. Biochem.}, 73:\penalty0 209--239, 2004.
\newblock \doi{10.1146/annurev.biochem.73.011303.073844}.

\bibitem[Marchetti et~al.(2013)Marchetti, Joanny, Ramaswamy, Liverpool, Prost,
  Rao, and Simha]{MJRL2013rmp}
M.~C. Marchetti, J.~F. Joanny, S.~Ramaswamy, T.~B. Liverpool, J.~Prost, M.~Rao,
  and R.~A. Simha.
\newblock Hydrodynamics of soft active matter.
\newblock \emph{Rev. Mod. Phys.}, 85:\penalty0 1143--1189, 2013.
\newblock \doi{10.1103/RevModPhys.85.1143}.

\bibitem[Jülicher et~al.(2007)Jülicher, Kruse, Prost, and Joanny]{JKPJ2007pr}
F.~Jülicher, K.~Kruse, J.~Prost, and J.~Joanny.
\newblock Active behavior of the cytoskeleton.
\newblock \emph{Phys. Rep.}, 449:\penalty0 3--28, 2007.
\newblock \doi{10.1016/j.physrep.2007.02.018}.

\bibitem[Bois et~al.(2011)Bois, J{\"u}licher, and Grill]{BoJG2011prl}
J.~S. Bois, F.~J{\"u}licher, and S.~W. Grill.
\newblock Pattern formation in active fluids.
\newblock \emph{Phys. Rev. Lett.}, 106:\penalty0 028103, 2011.
\newblock \doi{10.1103/PhysRevLett.106.028103}.

\bibitem[Yochelis et~al.(2016)Yochelis, Bar-On, and Gov]{YoBG2016pd}
A.~Yochelis, T.~Bar-On, and N.~S. Gov.
\newblock Reaction-diffusion-advection approach to spatially localized
  treadmilling aggregates of molecular motors.
\newblock \emph{Physica D}, 318:\penalty0 84--90, 2016.
\newblock \doi{10.1016/j.physd.2015.10.023}.

\bibitem[Shelley(2016)]{Shel2016arfm}
M.~J. Shelley.
\newblock The dynamics of microtubule/motor-protein assemblies in biology and
  physics.
\newblock \emph{Annu. Rev. Fluid Mech.}, 48:\penalty0 487--506, 2016.
\newblock \doi{10.1146/annurev-fluid-010814-013639}.

\bibitem[Bruinsma et~al.(2014)Bruinsma, Grosberg, Rabin, and
  Zidovska]{BGRZ2014bj}
R.~Bruinsma, A.~Grosberg, Y.~Rabin, and A.~Zidovska.
\newblock Chromatin hydrodynamics.
\newblock \emph{Biophys. J.}, 106:\penalty0 1871--1881, 2014.
\newblock \doi{10.1016/j.bpj.2014.03.038}.

\bibitem[Radszuweit et~al.(2014)Radszuweit, Engel, and B{\"a}r]{RaEB2014po}
M.~Radszuweit, H.~Engel, and M.~B{\"a}r.
\newblock An active poroelastic model for mechanochemical patterns in
  protoplasmic droplets of \textit{Physarum polycephalum}.
\newblock \emph{PLoS One}, 9:\penalty0 e99220, 2014.
\newblock \doi{10.1371/journal.pone.0099220}.

\bibitem[Kulawiak et~al.(2019)Kulawiak, Löber, Bär, and Engel]{KuJBE2019po}
D.~A. Kulawiak, J.~Löber, M.~Bär, and H.~Engel.
\newblock Active poroelastic two-phase model for the motion of physarum
  microplasmodia.
\newblock \emph{PLoS One}, 14:\penalty0 e0217447, 2019.
\newblock \doi{10.1371/journal.pone.0217447}.

\bibitem[Seminara et~al.(2012)Seminara, Angelini, Wilking, Vlamakis, Ebrahim,
  Kolter, Weitz, and Brenner]{SAWV2012pnasusa}
A.~Seminara, T.~E. Angelini, J.~N. Wilking, H.~Vlamakis, S.~Ebrahim, R.~Kolter,
  D.~A. Weitz, and M.~P. Brenner.
\newblock Osmotic spreading of \textit{{B}acillus subtilis} biofilms driven by
  an extracellular matrix.
\newblock \emph{Proc. Natl. Acad. Sci. U. S. A.}, 109:\penalty0 1116--1121,
  2012.
\newblock \doi{10.1073/pnas.1109261108}.

\bibitem[Trinschek et~al.(2017)Trinschek, John, Lecuyer, and
  Thiele]{TJLT2017prl}
S.~Trinschek, K.~John, S.~Lecuyer, and U.~Thiele.
\newblock Continuous vs. arrested spreading of biofilms at solid-gas interfaces
  - the role of surface forces.
\newblock \emph{Phys. Rev. Lett.}, 119:\penalty0 078003, 2017.
\newblock \doi{10.1103/PhysRevLett.119.078003}.

\bibitem[John and B{\"a}r(2005{\natexlab{a}})]{JoBa2005pb}
K.~John and M.~B{\"a}r.
\newblock Travelling lipid domains in a dynamic model for protein-induced
  pattern formation in biomembranes.
\newblock \emph{Phys. Biol.}, 2:\penalty0 123--132, 2005{\natexlab{a}}.
\newblock \doi{10.1088/1478-3975/2/2/005}.

\bibitem[John and B{\"a}r(2005{\natexlab{b}})]{JoBa2005prl}
K.~John and M.~B{\"a}r.
\newblock Alternative mechanisms of structuring biomembranes: self-assembly
  versus self-organization.
\newblock \emph{Phys. Rev. Lett.}, 95:\penalty0 198101, 2005{\natexlab{b}}.
\newblock \doi{10.1103/PhysRevLett.95.198101}.

\bibitem[Halatek and Frey(2018)]{HaFr2018np}
J.~Halatek and E.~Frey.
\newblock Rethinking pattern formation in reaction-diffusion systems.
\newblock \emph{Nature Phys.}, 14:\penalty0 507--514, 2018.
\newblock \doi{10.1038/s41567-017-0040-5}.

\bibitem[Halatek et~al.(2018)Halatek, Brauns, and Frey]{HaBF2018ptrsbs}
J.~Halatek, F.~Brauns, and E.~Frey.
\newblock Self-organization principles of intracellular pattern formation.
\newblock \emph{Philos. Trans. R. Soc. B-Biol. Sci.}, 373:\penalty0 20170107,
  2018.
\newblock \doi{10.1098/rstb.2017.0107}.

\bibitem[Banani et~al.(2017)Banani, Lee, Hyman, and Rosen]{BLHR2017nrmcb}
S.~F. Banani, H.~O. Lee, A.~A. Hyman, and M.~K. Rosen.
\newblock Biomolecular condensates: organizers of cellular biochemistry.
\newblock \emph{Nat. Rev. Mol. Cell Biol.}, 18:\penalty0 285--298, 2017.
\newblock \doi{10.1038/nrm.2017.7}.

\bibitem[Hyman et~al.(2014)Hyman, Weber, and Jülicher]{HyWJ2014arcdb}
A.~A. Hyman, C.~A. Weber, and F.~Jülicher.
\newblock Liquid-liquid phase separation in biology.
\newblock \emph{Annu. Rev. Cell Dev. Biol.}, 30:\penalty0 39--58, 2014.
\newblock \doi{10.1146/annurev-cellbio-100913-013325}.

\bibitem[Kirschbaum and Zwicker(2021)]{KiZw2021jotrsi}
J.~Kirschbaum and D.~Zwicker.
\newblock Controlling biomolecular condensates via chemical reactions.
\newblock \emph{J. R. Soc. Interface}, 18:\penalty0 20210255, 2021.
\newblock \doi{10.1098/rsif.2021.0255}.

\bibitem[Demarchi et~al.(2023)Demarchi, Goychuk, Maryshev, and
  Frey]{DGMF2023prl}
L.~Demarchi, A.~Goychuk, I.~Maryshev, and E.~Frey.
\newblock Enzyme-enriched condensates show self-propulsion, positioning, and
  coexistence.
\newblock \emph{Phys. Rev. Lett.}, 130:\penalty0 128401, 2023.
\newblock \doi{10.1103/PhysRevLett.130.128401}.

\bibitem[Goychuk et~al.(2024)Goychuk, Demarchi, Maryshev, and
  Frey]{GDMF2024prr}
A.~Goychuk, L.~Demarchi, I.~Maryshev, and E.~Frey.
\newblock Self-consistent sharp interface theory of active condensate dynamics.
\newblock \emph{Phys. Rev. Research}, 6:\penalty0 033082, 2024.
\newblock \doi{10.1103/physrevresearch.6.033082}.

\bibitem[Ismagilov et~al.(2002)Ismagilov, Schwartz, Bowden, and
  Whitesides]{ISBW2002acie}
R.~F. Ismagilov, A.~Schwartz, N.~Bowden, and G.~M. Whitesides.
\newblock Autonomous movement and self-assembly.
\newblock \emph{Angew. Chem.}, 41:\penalty0 652--654, 2002.
\newblock \doi{10.1002/1521-3773(20020215)41:4<652::aid-anie652>3.0.co;2-u}.

\bibitem[Paxton et~al.(2004)Paxton, Kistler, Olmeda, Sen, St.~Angelo, Cao,
  Mallouk, Lammert, and Crespi]{PKOS2004jotacs}
W.~F. Paxton, K.~C. Kistler, C.~C. Olmeda, A.~Sen, S.~K. St.~Angelo, Y.~Cao,
  T.~E. Mallouk, P.~E. Lammert, and V.~H. Crespi.
\newblock Catalytic nanomotors: Autonomous movement of striped nanorods.
\newblock \emph{J. Am. Chem. Soc.}, 126:\penalty0 13424--13431, 2004.
\newblock \doi{10.1021/ja047697z}.

\bibitem[Howse et~al.(2007)Howse, Jones, Ryan, Gough, Vafabakhsh, and
  Golestanian]{HJRG2007prla}
J.~R. Howse, R.~A.~L. Jones, A.~J. Ryan, T.~Gough, R.~Vafabakhsh, and
  R.~Golestanian.
\newblock Self-motile colloidal particles: From directed propulsion to random
  walk.
\newblock \emph{Phys. Rev. Lett.}, 99:\penalty0 048102, 2007.
\newblock \doi{10.1103/physrevlett.99.048102}.

\bibitem[Golestanian et~al.(2007)Golestanian, Liverpool, and
  Ajdari]{GoLA2007njop}
R.~Golestanian, T.~B. Liverpool, and A.~Ajdari.
\newblock Designing phoretic micro- and nano-swimmers.
\newblock \emph{New J. Phys.}, 9:\penalty0 126--126, 2007.
\newblock \doi{10.1088/1367-2630/9/5/126}.

\bibitem[Michelin and Lauga(2017)]{MiLa2017sr}
S.~Michelin and E.~Lauga.
\newblock Geometric tuning of self-propulsion for {J}anus catalytic particles.
\newblock \emph{Sci. Rep.}, 7, 2017.
\newblock \doi{10.1038/srep42264}.

\bibitem[Domingues Dos~Santos and Ondar{\c{c}}uhu(1995)]{DoOn1995prl}
F.~Domingues Dos~Santos and T.~Ondar{\c{c}}uhu.
\newblock Free-running droplets.
\newblock \emph{Phys. Rev. Lett.}, 75:\penalty0 2972--2975, 1995.
\newblock \doi{10.1103/PhysRevLett.75.2972}.

\bibitem[Lee and Laibinis(2000)]{LeLa2000jacs}
S.~W. Lee and P.~E. Laibinis.
\newblock Directed movement of liquids on patterned surfaces using noncovalent
  molecular adsorption.
\newblock \emph{J. Am. Chem. Soc.}, 122:\penalty0 5395--5396, 2000.
\newblock \doi{10.1021/ja994076a}.

\bibitem[Lee et~al.(2002)Lee, Kwok, and Laibinis]{LeKL2002pre}
S.~W. Lee, D.~Y. Kwok, and P.~E. Laibinis.
\newblock Chemical influences on adsorption-mediated self-propelled drop
  movement.
\newblock \emph{Phys. Rev. E}, 65:\penalty0 051602, 2002.
\newblock \doi{10.1103/PhysRevE.65.051602}.

\bibitem[Sumino et~al.(2005)Sumino, Kitahata, Yoshikawa, Nagayama, Nomura,
  Magome, and Mori]{SKYN2005pre}
Y.~Sumino, H.~Kitahata, K.~Yoshikawa, M.~Nagayama, S.~M. Nomura, N.~Magome, and
  Y.~Mori.
\newblock Chemosensitive running droplet.
\newblock \emph{Phys. Rev. E}, 72:\penalty0 041603, 2005.
\newblock \doi{10.1103/PhysRevE.72.041603}.

\bibitem[Thiele et~al.(2004)Thiele, John, and B{\"a}r]{ThJB2004prl}
U.~Thiele, K.~John, and M.~B{\"a}r.
\newblock Dynamical model for chemically driven running droplets.
\newblock \emph{Phys. Rev. Lett.}, 93:\penalty0 027802, 2004.
\newblock \doi{10.1103/PhysRevLett.93.027802}.

\bibitem[John et~al.(2005)John, B{\"a}r, and Thiele]{JoBT2005epje}
K.~John, M.~B{\"a}r, and U.~Thiele.
\newblock Self-propelled running droplets on solid substrates driven by
  chemical reactions.
\newblock \emph{Eur. Phys. J. E}, 18:\penalty0 183--199, 2005.
\newblock \doi{10.1140/epje/i2005-10039-1}.

\bibitem[Ziebert et~al.(2012)Ziebert, Swaminathan, and Aranson]{ZiSA2012jrsi}
F.~Ziebert, S.~Swaminathan, and I.~S. Aranson.
\newblock Model for self-polarization and motility of keratocyte fragments.
\newblock \emph{J. R. Soc. Interface}, 9:\penalty0 1084--1092, 2012.
\newblock \doi{10.1098/rsif.2011.0433}.

\bibitem[Tjhung et~al.(2015)Tjhung, Tiribocchi, Marenduzzo, and
  Cates]{TTMC2015nc}
E.~Tjhung, A.~Tiribocchi, D.~Marenduzzo, and M.~E. Cates.
\newblock A minimal physical model captures the shapes of crawling cells.
\newblock \emph{Nat. Commun.}, 6, 2015.
\newblock \doi{10.1038/ncomms6420}.

\bibitem[Trinschek et~al.(2020)Trinschek, Stegemerten, John, and
  Thiele]{TSJT2020prea}
S.~Trinschek, F.~Stegemerten, K.~John, and U.~Thiele.
\newblock Thin-film modeling of resting and moving active droplets.
\newblock \emph{Phys. Rev. E}, 101:\penalty0 062802, 2020.
\newblock \doi{10.1103/physreve.101.062802}.

\bibitem[Stegemerten et~al.(2022)Stegemerten, John, and Thiele]{StJT2022sm}
F.~Stegemerten, K.~John, and U.~Thiele.
\newblock Symmetry-breaking, motion and bistability of active drops through
  polarization-surface coupling.
\newblock \emph{Soft Matter}, 18:\penalty0 5823--5832, 2022.
\newblock \doi{10.1039/D2SM00648K}.

\bibitem[Maass et~al.(2016)Maass, Kr{\"u}ger, Herminghaus, and
  Bahr]{MKHB2016arcmp}
C.~C. Maass, C.~Kr{\"u}ger, S.~Herminghaus, and C.~Bahr.
\newblock Swimming droplets.
\newblock \emph{Annu. Rev. Condens. Matter Phys.}, 7:\penalty0 171--193, 2016.
\newblock \doi{10.1146/annurev-conmatphys-031115-011517}.

\bibitem[Michelin(2023)]{Mich2023arfm}
S.~Michelin.
\newblock Self-propulsion of chemically active droplets.
\newblock \emph{Annu. Rev. Fluid Mech.}, 55:\penalty0 77--101, 2023.
\newblock \doi{10.1146/annurev-fluid-120720-012204}.

\bibitem[Marincioni et~al.(2023)Marincioni, Nakashima, and Katsonis]{MaNK2023c}
B.~Marincioni, K.~K. Nakashima, and N.~Katsonis.
\newblock Motility of microscopic swimmers as protocells.
\newblock \emph{Chem}, 9:\penalty0 3030--3044, 2023.
\newblock \doi{10.1016/j.chempr.2023.10.007}.

\bibitem[Kitahata et~al.(2002)Kitahata, Aihara, Magome, and
  Yoshikawa]{KAMY2002jcp}
H.~Kitahata, R.~Aihara, N.~Magome, and K.~Yoshikawa.
\newblock Convective and periodic motion driven by a chemical wave.
\newblock \emph{J. Chem. Phys.}, 116:\penalty0 5666--5672, 2002.
\newblock \doi{10.1063/1.1456023}.

\bibitem[Kitahata et~al.(2011)Kitahata, Yoshinaga, Nagai, and
  Sumino]{KYNS2011pre}
H.~Kitahata, N.~Yoshinaga, K.~H. Nagai, and Y.~Sumino.
\newblock Spontaneous motion of a droplet coupled with a chemical wave.
\newblock \emph{Phys. Rev. E}, 84:\penalty0 015101, 2011.
\newblock \doi{10.1103/PhysRevE.84.015101}.

\bibitem[Suematsu et~al.(2016)Suematsu, Mori, Amemiya, and
  Nakata]{SMAN2016jpcl}
N.~J. Suematsu, Y.~Mori, T.~Amemiya, and S.~Nakata.
\newblock Oscillation of speed of a self-propelled {B}elousov-{Z}habotinsky
  droplet.
\newblock \emph{J. Phys. Chem. Lett.}, 7:\penalty0 3424--3428, 2016.
\newblock \doi{10.1021/acs.jpclett.6b01539}.

\bibitem[Szymanski et~al.(2013)Szymanski, Gorecki, and Hauser]{SzGH2013jpcc}
J.~Szymanski, J.~Gorecki, and M.~J.~B. Hauser.
\newblock Chemo-mechanical coupling in reactive droplets.
\newblock \emph{J. Phys. Chem. C}, 117:\penalty0 13080--13086, 2013.
\newblock \doi{10.1021/jp402308t}.

\bibitem[Lu et~al.(2013)Lu, Ren, Gao, Zhao, Wang, Yang, and
  Epstein]{LuRGZW2013cc}
X.~Lu, L.~Ren, Q.~Gao, Y.~Zhao, S.~Wang, J.~Yang, and I.~R. Epstein.
\newblock Photophobic and phototropic movement of a self-oscillating gel.
\newblock \emph{Chem. Commun.}, 49:\penalty0 7690, 2013.
\newblock \doi{10.1039/c3cc44480e}.

\bibitem[Ren et~al.(2020)Ren, Yuan, Gao, Teng, Wang, and Epstein]{RYG2020sa}
L.~Ren, L.~Yuan, Q.~Y. Gao, R.~Teng, J.~Wang, and I.~R. Epstein.
\newblock Chemomechanical origin of directed locomotion driven by internal
  chemical signals.
\newblock \emph{Sci. Adv.}, 6:\penalty0 eaaz9125, 2020.
\newblock \doi{10.1126/sciadv.aaz9125}.

\bibitem[Yashin and Balazs(2006)]{YaBa2006s}
V.~V. Yashin and A.~C. Balazs.
\newblock Pattern formation and shape changes in self-oscillating polymer gels.
\newblock \emph{Science}, 314:\penalty0 798--801, 2006.
\newblock \doi{10.1126/science.1132412}.

\bibitem[Paul and Joyce(2004)]{PaJo2004cocb}
N.~Paul and G.~F. Joyce.
\newblock Minimal self-replicating systems.
\newblock \emph{Curr. Opin. Chem. Biol.}, 8:\penalty0 634--639, 2004.
\newblock \doi{10.1016/j.cbpa.2004.09.005}.

\bibitem[Ruiz-Mirazo et~al.(2013)Ruiz-Mirazo, Briones, and de~la
  Escosura]{RuBE2013cr}
K.~Ruiz-Mirazo, C.~Briones, and A.~de~la Escosura.
\newblock Prebiotic systems chemistry: New perspectives for the origins of
  life.
\newblock \emph{Chem. Rev.}, 114:\penalty0 285--366, 2013.
\newblock \doi{10.1021/cr2004844}.

\bibitem[Adamski et~al.(2020)Adamski, Eleveld, Sood, Kun, Szil\'{a}gyi,
  Cz\'{a}r\'{a}n, Szathm\'{a}ry, and Otto]{AESK2020nrc}
P.~Adamski, M.~Eleveld, A.~Sood, A.~Kun, A.~Szil\'{a}gyi, T.~Cz\'{a}r\'{a}n,
  E.~Szathm\'{a}ry, and S.~Otto.
\newblock From self-replication to replicator systems en route to de novo life.
\newblock \emph{Nat. Rev. Chem.}, 4:\penalty0 386--403, 2020.
\newblock \doi{10.1038/s41570-020-0196-x}.

\bibitem[Rubinov et~al.(2009)Rubinov, Wagner, Rapaport, and
  Ashkenasy]{RWRA2009ac}
B.~Rubinov, N.~Wagner, H.~Rapaport, and G.~Ashkenasy.
\newblock Self‐replicating amphiphilic $\beta$‐sheet peptides.
\newblock \emph{Angew. Chem.}, 48:\penalty0 6683--6686, 2009.
\newblock \doi{10.1002/anie.200902790}.

\bibitem[Voss and Thiele(2024)]{VoTh2024jem}
F.~Voss and U.~Thiele.
\newblock Gradient dynamics approach to reactive thin-film hydrodynamics.
\newblock \emph{J. Eng. Math.}, 149:\penalty0 2, 2024.
\newblock \doi{10.1007/s10665-024-10402-x}.

\bibitem[Thiele et~al.(2018)Thiele, Snoeijer, Trinschek, and John]{TSTJ2018l}
U.~Thiele, J.~H. Snoeijer, S.~Trinschek, and K.~John.
\newblock Equilibrium contact angle and adsorption layer properties with
  surfactants.
\newblock \emph{Langmuir}, 34:\penalty0 7210--7221, 2018.
\newblock \doi{10.1021/acs.langmuir.8b00513}.
\newblock Also see Erratum: Langmuir, 35, 4788-4789 (2019),
  doi:10.1021/acs.langmuir.9b00616.

\bibitem[Bonn et~al.(2009)Bonn, Eggers, Indekeu, Meunier, and
  Rolley]{BEIM2009rmp}
D.~Bonn, J.~Eggers, J.~Indekeu, J.~Meunier, and E.~Rolley.
\newblock Wetting and spreading.
\newblock \emph{Rev. Mod. Phys.}, 81:\penalty0 739--805, 2009.
\newblock \doi{10.1103/RevModPhys.81.739}.

\bibitem[de~Gennes(1985)]{Genn1985rmp}
P.~G. de~Gennes.
\newblock Wetting: {S}tatics and dynamics.
\newblock \emph{Rev. Mod. Phys.}, 57:\penalty0 827--863, 1985.
\newblock \doi{10.1103/RevModPhys.57.827}.

\bibitem[Thiele(2010)]{Thie2010jpcm}
U.~Thiele.
\newblock Thin film evolution equations from (evaporating) dewetting liquid
  layers to epitaxial growth.
\newblock \emph{J. Phys. Condens. Matter}, 22:\penalty0 084019, 2010.
\newblock \doi{10.1088/0953-8984/22/8/084019}.

\bibitem[Thiele et~al.(2012)Thiele, Archer, and Plapp]{ThAP2012pf}
U.~Thiele, A.~J. Archer, and M.~Plapp.
\newblock Thermodynamically consistent description of the hydrodynamics of free
  surfaces covered by insoluble surfactants of high concentration.
\newblock \emph{Phys. Fluids}, 24:\penalty0 102107, 2012.
\newblock \doi{10.1063/1.4758476}.

\bibitem[Prigogine and Lefever(1968)]{PrLe1968jcp}
I.~Prigogine and R.~Lefever.
\newblock Symmetry breaking instabilities in dissipative systems. {II}.
\newblock \emph{J. Chem. Phys.}, 48:\penalty0 1695--1700, 1968.
\newblock \doi{10.1063/1.1668896}.

\bibitem[Nicolis(1999)]{Nicolis1999}
G.~Nicolis.
\newblock \emph{Introduction to Nonlinear Science}.
\newblock Cambridge University Press, Cambridge, 1999.
\newblock \doi{10.1017/CBO9781139170802}.

\bibitem[Satoh et~al.(2017)Satoh, Sogabe, Kayahara, Tanaka, Nagayama, and
  Nakata]{SSK2017sm}
Y.~Satoh, Y.~Sogabe, K.~Kayahara, S.~Tanaka, M.~Nagayama, and S.~Nakata.
\newblock Self-inverted reciprocation of an oil droplet on a surfactant
  solution.
\newblock \emph{Soft Matter}, 13:\penalty0 3422--3430, 2017.
\newblock \doi{10.1039/c7sm00252a}.

\bibitem[Tanaka et~al.(2021)Tanaka, Nakata, and Nagayama]{TaNN2021sm}
S.~Tanaka, S.~Nakata, and M.~Nagayama.
\newblock A surfactant reaction model for the reciprocating motion of a
  self-propelled droplet.
\newblock \emph{Soft Matter}, 17:\penalty0 388--396, 2021.
\newblock \doi{10.1039/d0sm01500h}.

\bibitem[Remlein et~al.(2025)Remlein, Esposito, and Avanzini]{ReEA2025jcp}
B.~Remlein, M.~Esposito, and F.~Avanzini.
\newblock What is a chemostat? insights from hybrid dynamics and stochastic
  thermodynamics.
\newblock \emph{J. Chem. Phys.}, 162:\penalty0 224113, 2025.
\newblock \doi{10.1063/5.0267465}.

\bibitem[Thiele et~al.(2016)Thiele, Archer, and Pismen]{ThAP2016prf}
U.~Thiele, A.~J. Archer, and L.~M. Pismen.
\newblock Gradient dynamics models for liquid films with soluble surfactant.
\newblock \emph{Phys. Rev. Fluids}, 1:\penalty0 083903, 2016.
\newblock \doi{10.1103/PhysRevFluids.1.083903}.

\bibitem[de~Groot and Mazur(1984)]{GrootMazur1984}
S.~R. de~Groot and P.~Mazur.
\newblock \emph{Non-equilibrium Thermodynamics}.
\newblock Dover publications, New York, 1984.

\bibitem[Onsager(1931{\natexlab{a}})]{Onsa1931prb}
L.~Onsager.
\newblock Reciprocal relations in irreversible processes. {I}.
\newblock \emph{Phys. Rev.}, 37:\penalty0 405--426, 1931{\natexlab{a}}.
\newblock \doi{10.1103/PhysRev.37.405}.

\bibitem[Onsager(1931{\natexlab{b}})]{Onsa1931pr}
L.~Onsager.
\newblock Reciprocal relations in irreversible processes. {II}.
\newblock \emph{Phys. Rev.}, 38:\penalty0 2265--2279, 1931{\natexlab{b}}.
\newblock \doi{10.1103/PhysRev.38.2265}.

\bibitem[Pereira et~al.(2007)Pereira, Trevelyan, Thiele, and
  Kalliadasis]{PTTK2007pf}
A.~Pereira, P.~M.~J. Trevelyan, U.~Thiele, and S.~Kalliadasis.
\newblock Dynamics of a horizontal thin liquid film in the presence of reactive
  surfactants.
\newblock \emph{Phys. Fluids}, 19:\penalty0 112102, 2007.
\newblock \doi{10.1063/1.2775938}.

\bibitem[Merkt et~al.(2005)Merkt, Pototsky, Bestehorn, and Thiele]{MPBT2005pf}
D.~Merkt, A.~Pototsky, M.~Bestehorn, and U.~Thiele.
\newblock Long-wave theory of bounded two-layer films with a free liquid-liquid
  interface: {S}hort- and long-time evolution.
\newblock \emph{Phys. Fluids}, 17:\penalty0 064104, 2005.
\newblock \doi{10.1063/1.1935487}.

\bibitem[Weber et~al.(2019)Weber, Zwicker, J{\"{u}}licher, and
  Lee]{WZJL2019rpp}
C.~A. Weber, D.~Zwicker, F.~J{\"{u}}licher, and C.~F. Lee.
\newblock Physics of active emulsions.
\newblock \emph{Rep. Prog. Phys.}, 82:\penalty0 064601, 2019.
\newblock \doi{10.1088/1361-6633/ab052b}.

\bibitem[Zwicker(2022)]{Zwic2022cocis}
D.~Zwicker.
\newblock The intertwined physics of active chemical reactions and phase
  separation.
\newblock \emph{Curr. Opin. Colloid Interface Sci.}, 61:\penalty0 101606, 2022.
\newblock \doi{10.1016/j.cocis.2022.101606}.

\bibitem[Avanzini et~al.(2024)Avanzini, Aslyamov, Fodor, and
  Esposito]{AvAFE2024}
F.~Avanzini, T.~Aslyamov, {\'{E}}.~Fodor, and M.~Esposito.
\newblock Nonequilibrium thermodynamics of non-ideal reaction–diffusion
  systems: Implications for active self-organization.
\newblock \emph{J. Chem. Phys.}, 161:\penalty0 174108, 2024.
\newblock \doi{10.1063/5.0231520}.

\bibitem[Heil and Hazel(2006)]{HeHa2006}
M.~Heil and A.~L. Hazel.
\newblock Oomph-lib - an object-oriented multi-physics finite-element library.
\newblock In H.-J. Bungartz and M.~Sch{\"a}fer, editors, \emph{Fluid-Structure
  Interaction: Modelling, Simulation, Optimisation}, pages 19--49. Springer,
  Berlin, Heidelberg, 2006.
\newblock \doi{10.1007/3-540-34596-5_2}.

\bibitem[Frohoff-H{\"u}lsmann and Thiele(2023)]{FrTh2023prl}
T.~Frohoff-H{\"u}lsmann and U.~Thiele.
\newblock Nonreciprocal {C}ahn-{H}illiard model emerges as a universal
  amplitude equation.
\newblock \emph{Phys. Rev. Lett.}, 131:\penalty0 107201, 2023.
\newblock \doi{10.1103/PhysRevLett.131.107201}.

\bibitem[Krauskopf et~al.(2007)Krauskopf, Osinga, and
  Galan-Vioque]{KrauskopfOsingaGalan-Vioque2007}
B.~Krauskopf, H.~M. Osinga, and J.~Galan-Vioque, editors.
\newblock \emph{Numerical Continuation Methods for Dynamical Systems}.
\newblock Springer, Dordrecht, 2007.
\newblock \doi{10.1007/978-1-4020-6356-5}.

\bibitem[Dijkstra et~al.(2014)Dijkstra, Wubs, Cliffe, Doedel, Dragomirescu,
  Eckhardt, Gelfgat, Hazel, Lucarini, Salinger, Phipps, Sanchez-Umbria,
  Schuttelaars, Tuckerman, and Thiele]{DWCD2014ccp}
H.~A. Dijkstra, F.~W. Wubs, A.~K. Cliffe, E.~Doedel, I.~F. Dragomirescu,
  B.~Eckhardt, A.~Y. Gelfgat, A.~Hazel, V.~Lucarini, A.~G. Salinger, E.~T.
  Phipps, J.~Sanchez-Umbria, H.~Schuttelaars, L.~S. Tuckerman, and U.~Thiele.
\newblock Numerical bifurcation methods and their application to fluid
  dynamics: {A}nalysis beyond simulation.
\newblock \emph{Commun. Comput. Phys.}, 15:\penalty0 1--45, 2014.
\newblock \doi{10.4208/cicp.240912.180613a}.

\bibitem[Engelnkemper et~al.(2019)Engelnkemper, Gurevich, Uecker, Wetzel, and
  Thiele]{EGUW2019springer}
S.~Engelnkemper, S.~V. Gurevich, H.~Uecker, D.~Wetzel, and U.~Thiele.
\newblock Continuation for thin film hydrodynamics and related scalar problems.
\newblock In A.~Gelfgat, editor, \emph{Computational Modeling of Bifurcations
  and Instabilities in Fluid Mechanics}, Computational Methods in Applied
  Sciences, vol 50, pages 459--501. Springer, Cham, 2019.
\newblock \doi{10.1007/978-3-319-91494-7_13}.

\bibitem[Uecker(2021)]{Uecker2021}
H.~Uecker.
\newblock \emph{Numerical Continuation and Bifurcation in Nonlinear PDEs}.
\newblock SIAM, 2021.
\newblock \doi{10.1137/1.9781611976618}.

\bibitem[Gambaudo et~al.(1988)Gambaudo, Glendinning, and Tresser]{GaGT1988n}
J.~M. Gambaudo, P.~Glendinning, and C.~Tresser.
\newblock The gluing bifurcation: I. {S}ymbolic dynamics of closed curves.
\newblock \emph{Nonlinearity}, 1:\penalty0 203--214, 1988.
\newblock \doi{10.1088/0951-7715/1/1/008}.

\bibitem[Paz{\'o} and P{\'e}rez-Mu{\~n}uzuri(2001)]{PaPe2001pre}
D.~Paz{\'o} and V.~P{\'e}rez-Mu{\~n}uzuri.
\newblock Onset of wave fronts in a discrete bistable medium.
\newblock \emph{Phys. Rev. E}, 64:\penalty0 065203, 2001.
\newblock \doi{10.1103/PhysRevE.64.065203}.

\bibitem[Herrero et~al.(1998)Herrero, Farjas, Pons, Pi, and
  Orriols]{HFPP1998pre}
R.~Herrero, J.~Farjas, R.~Pons, F.~Pi, and G.~Orriols.
\newblock Gluing bifurcations in optothermal nonlinear devices.
\newblock \emph{Phys. Rev. E}, 57:\penalty0 5366--5377, 1998.
\newblock \doi{10.1103/PhysRevE.57.5366}.

\bibitem[Lopez and Marques(2000)]{LoMa2000prl}
J.~M. Lopez and F.~Marques.
\newblock Dynamics of three-tori in a periodically forced {Navier-Stokes} flow.
\newblock \emph{Phys. Rev. Lett.}, 85:\penalty0 972--975, 2000.
\newblock \doi{10.1103/physrevlett.85.972}.

\bibitem[Demeter and Kramer(1999)]{DeKr1999prl}
G.~Demeter and L.~Kramer.
\newblock Transition to chaos via gluing bifurcations in optically excited
  nematic liquid crystals.
\newblock \emph{Phys. Rev. Lett.}, 83:\penalty0 4744--4747, 1999.
\newblock \doi{10.1103/PhysRevLett.83.4744}.

\bibitem[Kuramoto and Koga(1982)]{KuKo1982pla}
Y.~Kuramoto and S.~Koga.
\newblock Anomalous period-doubling bifurcations leading to chemical
  turbulence.
\newblock \emph{Phys. Lett. A}, 92:\penalty0 1--4, 1982.
\newblock \doi{10.1016/0375-9601(82)90725-3}.

\bibitem[Abshagen et~al.(2001)Abshagen, Pfister, and Mullin]{AbPM2001prl}
J.~Abshagen, G.~Pfister, and T.~Mullin.
\newblock Gluing bifurcations in a dynamically complicated extended flow.
\newblock \emph{Phys. Rev. Lett.}, 87:\penalty0 224501, 2001.
\newblock \doi{10.1103/physrevlett.87.224501}.

\bibitem[Tarama and Ohta(2016)]{TaOh2016el}
M.~Tarama and T.~Ohta.
\newblock Reciprocating motion of active deformable particles.
\newblock \emph{Europhys. Lett.}, 114:\penalty0 30002, 2016.
\newblock \doi{10.1209/0295-5075/114/30002}.

\bibitem[Koper(1995)]{Ko1995pd}
M.~T.~M. Koper.
\newblock Bifurcations of mixed-mode oscillations in a three-variable
  autonomous van der {P}ol-{D}uffing model with a cross-shaped phase diagram.
\newblock \emph{Physica D}, 80:\penalty0 72--94, 1995.
\newblock \doi{10.1016/0167-2789(95)90061-6}.

\bibitem[Shilnikov et~al.(2001)Shilnikov, Shilnikov, Turaev, and
  Chua]{ShilnikovShilnikovTuraevChua2001}
L.~P. Shilnikov, A.~L. Shilnikov, D.~V. Turaev, and L.~O. Chua.
\newblock \emph{Methods of Qualitative Theory in Nonlinear Dynamics: (Part
  II)}.
\newblock World Scientific, 2001.
\newblock \doi{10.1142/4221}.

\bibitem[Glendinning and Sparrow(1984)]{GlSp1984jsp}
P.~Glendinning and C.~Sparrow.
\newblock Local and global behavior near homoclinic orbits.
\newblock \emph{J. Stat. Phys.}, 35:\penalty0 645--696, 1984.
\newblock \doi{10.1007/bf01010828}.

\bibitem[Arneodo et~al.(1980)Arneodo, Coullet, and Tresser]{ArCT1980pla}
A.~Arneodo, P.~Coullet, and C.~Tresser.
\newblock Occurence of strange attractors in three-dimensional volterra
  equations.
\newblock \emph{Phys. Lett. A}, 79:\penalty0 259--263, 1980.
\newblock \doi{10.1016/0375-9601(80)90342-4}.

\bibitem[Moore et~al.(1983)Moore, Toomre, Knobloch, and Weiss]{MTKW1983n}
D.~R. Moore, J.~Toomre, E.~Knobloch, and N.~O. Weiss.
\newblock Period doubling and chaos in partial differential equations for
  thermosolutal convection.
\newblock \emph{Nature}, 303:\penalty0 663--667, 1983.
\newblock \doi{10.1038/303663a0}.

\bibitem[Knobloch et~al.(1986)Knobloch, Moore, Toomre, and Weiss]{KMTW1986jfm}
E.~Knobloch, D.~R. Moore, J.~Toomre, and N.~O. Weiss.
\newblock Transitions to chaos in two-dimensional double-diffusive convection.
\newblock \emph{J. Fluid Mech.}, 166:\penalty0 409, 1986.
\newblock \doi{10.1017/s0022112086000216}.

\bibitem[Knobloch et~al.(1999)Knobloch, Landsberg, and Moehlis]{KnLM1999pla}
E.~Knobloch, A.~S. Landsberg, and J.~Moehlis.
\newblock Chaotic direction-reversing waves.
\newblock \emph{Phys. Lett. A}, 255:\penalty0 287--293, 1999.
\newblock \doi{10.1016/S0375-9601(99)00200-5}.

\bibitem[Koper and Gaspard(1991)]{KoGa1991jpc}
M.~T.~M. Koper and P.~Gaspard.
\newblock Mixed-mode and chaotic oscillations in a simple model of an
  electrochemical oscillator.
\newblock \emph{J, Phys. Chem.}, 95:\penalty0 4945--4947, 1991.
\newblock \doi{10.1021/j100166a009}.

\bibitem[Koper and Gaspard(1992)]{KoGa1992jcp}
M.~T.~M. Koper and P.~Gaspard.
\newblock The modeling of mixed-mode and chaotic oscillations in
  electrochemical systems.
\newblock \emph{J. Chem. Phys.}, 96:\penalty0 7797--7813, 1992.
\newblock \doi{10.1063/1.462377}.

\bibitem[Guckenheimer and Lizarraga(2015)]{GuLi2015sjads}
J.~Guckenheimer and I.~Lizarraga.
\newblock Shilnikov homoclinic bifurcation of mixed-mode oscillations.
\newblock \emph{SIAM J. Appl. Dyn. Syst.}, 14:\penalty0 764--786, 2015.
\newblock \doi{10.1137/140972007}.

\bibitem[Glendinning and Sparrow(1986)]{GlSp1986jsp}
P.~Glendinning and C.~Sparrow.
\newblock T-points - a codimension 2 heteroclinic bifurcation.
\newblock \emph{J. Stat. Phys.}, 43:\penalty0 479--488, 1986.
\newblock \doi{10.1007/BF01020649}.

\bibitem[Bykov(1993)]{Byko1993pd}
V.~Bykov.
\newblock The bifurcations of separatrix contours and chaos.
\newblock \emph{Physica D}, 62:\penalty0 290--299, 1993.
\newblock \doi{10.1016/0167-2789(93)90288-c}.

\bibitem[Hirschberg and Knobloch(1993)]{HiKn1993pd}
P.~Hirschberg and E.~Knobloch.
\newblock Šil’nikov-{H}opf bifurcation.
\newblock \emph{Physica D}, 62:\penalty0 202--216, 1993.
\newblock \doi{10.1016/0167-2789(93)90282-6}.

\bibitem[Algaba et~al.(2015)Algaba, Fernández-Sánchez, Merino, and
  Rodríguez-Luis]{AlFeMeRo2015cnsms}
A.~Algaba, F.~Fernández-Sánchez, M.~Merino, and A.~Rodríguez-Luis.
\newblock {Analysis of the T-point-Hopf bifurcation in the Lorenz system}.
\newblock \emph{Commun. Nonlinear Sci. Numer. Simul.}, 22:\penalty0 676--691,
  2015.
\newblock \doi{10.1016/j.cnsns.2014.09.025}.

\bibitem[Champneys et~al.(2007)Champneys, Kirk, Knobloch, Oldeman, and
  Sneyd]{CKKO2007sjoads}
A.~R. Champneys, V.~Kirk, E.~Knobloch, B.~E. Oldeman, and J.~Sneyd.
\newblock When {Shil’nikov} meets {Hopf} in excitable systems.
\newblock \emph{SIAM J. Appl. Dyn. Syst.}, 6:\penalty0 663--693, 2007.
\newblock \doi{10.1137/070682654}.

\bibitem[del Castillo-Negrete(2008)]{delC2008}
D.~del Castillo-Negrete.
\newblock \emph{Fractional Diffusion Models of Anomalous Transport}, chapter~6,
  pages 163--212.
\newblock John Wiley \& Sons, Ltd, 2008.
\newblock \doi{10.1002/9783527622979.ch6}.

\bibitem[Hartmann et~al.(2023)Hartmann, Diddens, Jalaal, and
  Thiele]{HDJT2023jfm}
S.~Hartmann, C.~Diddens, M.~Jalaal, and U.~Thiele.
\newblock Sessile drop evaporation in a gap - crossover between
  diffusion-limited and phase transition-limited regime.
\newblock \emph{J. Fluid Mech.}, 960:\penalty0 A32, 2023.
\newblock \doi{10.1017/jfm.2023.176}.

\bibitem[Izhikevich(2000)]{IZHI2000ijobac}
E.~M. Izhikevich.
\newblock Neural excitability, spiking and bursting.
\newblock \emph{Int. J. Bifurcation Chaos}, 10:\penalty0 1171--1266, 2000.
\newblock \doi{10.1142/s0218127400000840}.

\bibitem[Dullweber et~al.(2025)Dullweber, Belousov, Autorino, Petridou, and
  Erzberger]{DuBAPE2025prl}
T.~Dullweber, R.~Belousov, C.~Autorino, N.~I. Petridou, and A.~Erzberger.
\newblock Shape switching and tunable oscillations of adaptive droplets.
\newblock \emph{Phys. Rev. Lett.}, 2025.
\newblock \doi{10.1103/1cq4-x499}.
\newblock (at press).

\bibitem[Ron et~al.(2020)Ron, Monzo, Gauthier, Voituriez, and Gov]{RMGV2020prr}
J.~E. Ron, P.~Monzo, N.~C. Gauthier, R.~Voituriez, and N.~S. Gov.
\newblock One-dimensional cell motility patterns.
\newblock \emph{Phys. Rev. Research}, 2:\penalty0 033237, 2020.
\newblock \doi{10.1103/physrevresearch.2.033237}.

\bibitem[Yochelis et~al.(2022)Yochelis, Flemming, and Beta]{YoFB2022prl}
A.~Yochelis, S.~Flemming, and C.~Beta.
\newblock Versatile patterns in the actin cortex of motile cells:
  Self-organized pulses can coexist with macropinocytic ring-shaped waves.
\newblock \emph{Phys. Rev. Lett.}, 129:\penalty0 088101, 2022.
\newblock \doi{10.1103/physrevlett.129.088101}.

\bibitem[Hughes et~al.(2025)Hughes, Modai, Edelstein-Keshet, and
  Yochelis]{HuMEY2024}
J.~M. Hughes, S.~Modai, L.~Edelstein-Keshet, and A.~Yochelis.
\newblock Bistability of travelling waves and wave-pinning states in a
  mass-conserved reaction-diffusion system: From bifurcations to implications
  for actin waves.
\newblock \emph{arXiv}, 2025.
\newblock \doi{10.48550/arXiv.2410.12213}.

\bibitem[Ophaus et~al.(2020)Ophaus, Kirchner, Gurevich, and Thiele]{OKGT2020c}
L.~Ophaus, J.~Kirchner, S.~V. Gurevich, and U.~Thiele.
\newblock Phase-field-crystal description of active crystallites: {E}lastic and
  inelastic collisions.
\newblock \emph{Chaos}, 30:\penalty0 123149, 2020.
\newblock \doi{10.1063/5.0019426}.

\bibitem[Landsberg and Knobloch(1991)]{LaKn1991pla}
A.~S. Landsberg and E.~Knobloch.
\newblock Direction-reversing traveling waves.
\newblock \emph{Phys. Lett. A}, 159:\penalty0 17--20, 1991.
\newblock \doi{10.1016/0375-9601(91)90155-2}.

\bibitem[Magome and Yoshikawa(1996)]{MaYo1996jpc}
N.~Magome and K.~Yoshikawa.
\newblock Nonlinear oscillation and ameba-like motion in an oil/water system.
\newblock \emph{J. Phys. Chem.}, 100:\penalty0 19102--19105, 1996.
\newblock \doi{10.1021/jp9616876}.

\bibitem[Tanaka et~al.(2015)Tanaka, Sogabe, and Nakata]{TaSN2015pre}
S.~Tanaka, Y.~Sogabe, and S.~Nakata.
\newblock Spontaneous change in trajectory patterns of a self-propelled oil
  droplet at the air-surfactant solution interface.
\newblock \emph{Phys. Rev. E}, 91:\penalty0 032406, 2015.
\newblock \doi{10.1103/PhysRevE.91.032406}.

\bibitem[Dreher et~al.(2014)Dreher, Aranson, and Kruse]{DrAK2014njp}
A.~Dreher, I.~S. Aranson, and K.~Kruse.
\newblock Spiral actin-polymerization waves can generate amoeboidal cell
  crawling.
\newblock \emph{New J. Phys.}, 16:\penalty0 055007, 2014.
\newblock \doi{10.1088/1367-2630/16/5/055007}.

\bibitem[Rodiek et~al.(2015)Rodiek, Takagi, Ueda, and Hauser]{RoTUH2015ebj}
B.~Rodiek, S.~Takagi, T.~Ueda, and M.~J.~B. Hauser.
\newblock Patterns of cell thickness oscillations during directional migration
  of \textit{Physarum polycephalum}.
\newblock \emph{Eur. Biophys. J.}, 44:\penalty0 349--358, 2015.
\newblock \doi{10.1007/s00249-015-1028-7}.

\bibitem[Lewis et~al.(2015)Lewis, Zhang, Guy, and del Álamo]{LeZGA2015jrsi}
O.~L. Lewis, S.~Zhang, R.~D. Guy, and J.~C. del Álamo.
\newblock Coordination of contractility, adhesion and flow in migrating
  \textit{Physarum} amoebae.
\newblock \emph{J. R. Soc. Interface}, 12:\penalty0 20141359, 2015.
\newblock \doi{10.1098/rsif.2014.1359}.

\bibitem[Roosen-Runge et~al.(2011)Roosen-Runge, Hennig, Zhang, Jacobs, Sztucki,
  Schober, Seydel, and Schreiber]{RHZJ2011pnas}
F.~Roosen-Runge, M.~Hennig, F.~Zhang, R.~M.~J. Jacobs, M.~Sztucki, H.~Schober,
  T.~Seydel, and F.~Schreiber.
\newblock Protein self-diffusion in crowded solutions.
\newblock \emph{Proc. Natl. Acad. Sci. U.S.A.}, 108:\penalty0 11815--11820,
  2011.
\newblock \doi{10.1073/pnas.1107287108}.

\bibitem[Craster and Matar(2009)]{CrMa2009rmp}
R.~V. Craster and O.~K. Matar.
\newblock Dynamics and stability of thin liquid films.
\newblock \emph{Rev. Mod. Phys.}, 81:\penalty0 1131--1198, 2009.
\newblock \doi{10.1103/RevModPhys.81.1131}.

\bibitem[Trinschek et~al.(2018)Trinschek, John, and Thiele]{TrJT2018sm}
S.~Trinschek, K.~John, and U.~Thiele.
\newblock Modelling of surfactant-driven front instabilities in spreading
  bacterial colonies.
\newblock \emph{Soft Matter}, 14:\penalty0 4464--4476, 2018.
\newblock \doi{10.1039/c8sm00422f}.

\bibitem[Sakata and Berg(1969)]{SaBe1969iecf}
E.~K. Sakata and J.~C. Berg.
\newblock Surface diffusion in monolayers.
\newblock \emph{Ind. Eng. Chem. Fundam.}, 8:\penalty0 570--575, 1969.
\newblock \doi{10.1021/i160031a033}.

\bibitem[Nicolas(2003)]{Ni2003bj}
J.~Nicolas.
\newblock Molecular dynamics simulation of surfactin molecules at the
  water-hexane interface.
\newblock \emph{Biophys. J.}, 85:\penalty0 1377--1391, 2003.
\newblock \doi{10.1016/s0006-3495(03)74571-8}.

\bibitem[Bai and Breen(2008)]{BaBre2008}
L.~Bai and D.~Breen.
\newblock Calculating center of mass in an unbounded 2d environment.
\newblock \emph{J. Graph. Tools}, 13:\penalty0 53--60, 2008.
\newblock \doi{10.1080/2151237x.2008.10129266}.

\bibitem[Voss and Thiele(2025)]{VoTh2025zn}
F.~Voss and U.~Thiele.
\newblock {Data and Code Supplement - "Chemomechanical motility modes of
  partially wetting liquid droplets"}.
\newblock Zenodo, 2025.
\newblock \doi{10.5281/zenodo.15880245}.

\end{thebibliography}
\end{document}